\documentclass[a4paper,11pt]{article}

\pdfoutput=1

\usepackage{amsmath}
\usepackage{amsfonts}
\usepackage{amssymb}
\usepackage{comment}
\usepackage[colorlinks,citecolor=blue,urlcolor=blue,bookmarks=false,hypertexnames=true]{hyperref}
\usepackage{cite}

\usepackage{graphicx, physics, subcaption}
\usepackage{verbatim }

\numberwithin{equation}{section}

\usepackage{float} % In the preamble

\usepackage{color}

\usepackage{booktabs}

\usepackage{amsmath, amssymb, graphics, epsfig, graphicx}
\usepackage{epsf}
\usepackage{epstopdf}
\usepackage {amssymb}
\newcommand{\nc}{\newcommand}
\usepackage{amsmath}
\usepackage{amsthm}

\newtheorem{definition}{Definition}

\newcommand{\calR}{{\cal{R}}}

\def\bfk{{\bf k}}

\def\bfx{{\bf x}}

\nc{\cN}{ {\cal{N}} }

\nc{\ba}{\begin{eqnarray}}
\nc{\ea}{\end{eqnarray}}

\usepackage{amsmath,amssymb}
\usepackage{tikz}
\usepackage{float}

\tikzset{
  dnvertex/.style={
    circle,
    draw,
    fill=white,
    minimum size=6mm,
    inner sep=0pt,
    font=\small
  },
  dnedge/.style={line width=0.7pt},
  dnmomentum/.style={
    font=\scriptsize,
    fill=white,
    inner sep=1pt
  }
}

\usepackage{amsmath,amssymb,booktabs,tikz}
\begin{document}

%%%%%%%%%%%%%%%%%%%%%%%%%%%%%%%%%%%
\vspace{5mm}
\vspace{0.5cm}
\begin{center}

\def\thefootnote{\fnsymbol{footnote}}

{\bf\large {{General $\delta N$ formalism in Single Field Inflation and Graph Theory } }}
\\[0.5cm]

{  Amin Nassiri-Rad$^a$ $\footnote{amin.nassiriraad@kntu.ac.ir}$, Hassan Firouzjahi$^b$ $\footnote{firouz@ipm.ir}$ }   \\[0.5cm]
\small \textit{$^a$ Department of physics, K.N Toosi University of Technology, P.O. Box 15875-4416, Tehran, Iran}
\\{\small \textit{$^b$ School of Astronomy, Institute for Research in Fundamental Sciences (IPM) \\ P.~O.~Box 19395-5746, Tehran, Iran}
} 

\end{center}

\vspace{.8cm}

\hrule 
\vspace{0.3cm}

%%%%%%%%%%%%%%%%%%%%%%%%%%%%%%%%%%%%%%%%%%%%%%%%%%%%%

\begin{abstract}
 
 We study higher order correlation functions in single field inflation using the $\delta N$ formalism and methods from graph theory. By representing connected tree level contributions as graphs, we systematically classify their independent topologies, momentum shapes and amplitudes. We show that the amplitudes of arbitrary trees factorize into products of primary amplitudes, 
 providing a general framework for relations among higher order correlators, including the Suyama-Yamaguchi equality. 
As the number of external legs increases, the number of distinct tree topologies grows exponentially, whereas the number of algebraically distinct amplitude products grows considerably more slowly. 
As an application, we consider ultra-slow-roll inflation followed by a transition to a slow-roll attractor. For transitions without a potential step, we derive an explicit all-order expression for the vertex amplitudes and establish their positivity and dependence on the sharpness of the transition. When a potential step is included, we prove that all nonlinear vertex amplitudes remain positive for upward steps, whereas downward steps allow cancellations at specific parameter values. These cancellations can eliminate entire families of tree contributions, including configurations in which the leading local bispectrum and trispectrum vanish simultaneously while 
 higher order correlations remain nonzero.

\end{abstract}
\vspace{0.5cm} \hrule
\def\thefootnote{\arabic{footnote}}
\setcounter{footnote}{0}
\newpage
%%%%%%%%%%%%%%%%%%%%%%%%%%%%%%%%%%%%%%%%%%%%%%%%%%%%%

%%%%%%%%%%%%%%%%%%%%%%%%%%%%%%%%%%%%%%
\section{Introduction}
Cosmological perturbation theory is the fundamental tool to study cosmological observables such as the cosmic microwave 
background (CMB) temperature anisotropies and the power spectrum or bispectrum of primordial curvature perturbations \cite{Kodama:1984ziu, Mukhanov:1990me, Weinberg:2008zzc, Baumann:2009ds}. The general view  is that these perturbations are generated from quantum fluctuations during inflation in early universe which eventually provide the seeds of CMB anisotropies and large scale structure. There are two main approaches to study primordial perturbations within cosmological perturbation theory, namely the standard quantum field theory approach and the alternative $\delta N$ formalism. In the standard QFT approach one promotes the cosmological observables, such as the curvature perturbations on comoving slice $\calR$  or the Bardeen potential, into quantum operators. These quantum operators are further endowed  with appropriate vacuum (initial conditions) deep inside the horizon such as the Baunch-Davies (Minkowski) vacuum and with the corresponding creation and annihilation operators. One then look at the evolution of the corresponding mode functions and calculate the expectation values such the two-point functions (power spectrum) and three-point functions (bispectrum) etc at the surface of end of inflation. This is achieved within the machinery of QFT in-in formalism \cite{Weinberg:2005vy, Maldacena:2002vr, Adshead:2009cb, Chen:2009zp}
where all information is encoded in the corresponding interaction Hamiltonian. 

On the other hand, in the alternative approach based on $\delta N$ formalism one  employs the separate universe approach in which perturbations on superhorizon scales follow the background trajectories \cite{Sasaki:1995aw, Sasaki:1998ug,  Wands:2000dp, Lyth:2004gb, Lyth:2005fi, Abolhasani:2019cqw}. Different patches of Hubble size $H^{-1}$ are viewed as independent FLRW backgrounds which are slightly different from each other by their initial conditions which are induced by the corresponding large scale perturbations. To calculate curvature perturbations, one only has to read the differences in the background e-fold number between an initial hypersurface and a final hypersurface. The initial hypersurface is spatially flat while the final surface is a surface of constant energy. The advantage of $\delta N$ formalism is that one only needs to follow the background trajectory and calculate the number of e-folds $N$ as a function of background fields and their velocities. Another important advantage of $\delta N$ approach is that it is a non-linear formalism so it does not need the perturbative expansion as is required in the standard QFT approach. Correspondingly, $\delta N$ is a powerful tool in calculating non-Gaussianities and the full probability density function for primordial black holes (PBHs) formation \cite{Hooshangi:2021ubn, Cai:2021zsp, Cai:2022erk, Kawaguchi:2023mgk}. $\delta N$ formalism has been employed extensively in literature to calculate bispectrum and trispectrum in single field and multiple fields setups, for an incomplete list see \cite{Lyth:2005qk, Sasaki:2008uc, Naruko:2008sq, Wands:2007bd, Wands:2010af, Cruces:2025typ, Pi:2022ysn, Cai:2018dkf,  Chen:2013eea, Namjoo:2012aa, Abolhasani:2012px,  Abolhasani:2018gyz, Talebian-Ashkezari:2016llx, Abolhasani:2013zya, Tanaka:2010km, Tanaka:2024mzw,Tanaka:2023gul, Tanaka:2021dww, Cruces:2022dom, Firouzjahi:2023ahg}. In addition, stochastic $\delta N$ formalism proved very fruitful to study cosmological correlators in the large diffusion limit within the context of stochastic inflation \cite{Fujita:2013cna, Fujita:2014tja, Vennin:2015hra,Vennin:2016wnk, Assadullahi:2016gkk,Grain:2017dqa, Noorbala:2018zlv,Firouzjahi:2018vet, Firouzjahi:2020jrj, Ahmadi:2022lsm, Noorbala:2024fim, Ahmadi:2025oon, Cruces:2018cvq, Miyamoto:2026sbo, Murata:2026yqb, Miyamoto:2025qqm, Murata:2025onc}.

In this work we take a different approach and look at the $\delta N$ formalism and the corresponding analysis of cosmological correlators from the graph theory point of view. It turns out that general and powerful mathematical conclusions can be established for the $n$-point correlation functions of comoving curvature perturbation in single field setup within the context of $\delta N$ formalism which can be nicely represented via graph theory.

%%%%%%%%%%%%%%%%%%%%%%%%%%%%%%%%%%%%%%
\section{Correlation Functions  }
\label{review}

Our goal in this section is to express the $n$-point correlation functions of curvature perturbations in single field inflation 
using $\delta N$ formalism. Here we present the shapes of correlation functions for the simple  cases of power spectrum ($n=2$), bispectrum ($n=3$) and trispectrum ($n=4$).  After this review we set the stage for the general $n$-point correlation function.  

The background is given by a flat FLRW metric,
\ba
ds^2= - dt^2+ a(t)^2 d \bfx^2 \, ,
\ea
in which $t$ is the cosmic time and $a(t)$ is the scale factor. Related to the scale factor, the Hubble expansion parameter is defined via $H\equiv \dot a/a$ in which a dot indicates the derivative with respect to $t$. We assume that inflation is driven by a single scalar field, the inflaton field $\phi$, with the potential $V(\phi)$. It is assumed that during inflation the energy density responsible for the expansion is predominantly provided by the potential and the inflaton kinetic energy is subleading. Defining the first slow-roll (SR) parameter $\epsilon \equiv -\dot H/H^2$, this assumption corresponds to 
$\epsilon \ll 1$. However, we do not assume that the system is always in  the SR regime. Indeed, there can be periods during inflation in which the second SR parameter $\eta= \dot \epsilon/H \epsilon$ may not be small such as in the ultra slow-roll (USR) setup. However, the assumption $\epsilon \ll 1$ ensures that we have  a near exponential  expansion for the scale factor $a(t) \sim e^{H t}$ with $H$ being nearly constant during inflation. 

In our analysis below, it is more useful to use the number of e-fold $N$ as the clock which is related to the cosmic time via $dN= H(t) dt$.  One may chose the convention that $N=0$ at the start of inflation (or when the CMB observables leave the Hubble horizon) while the time of end of inflation is denoted by $N=N_f$. To solve the flatness and the horizon problems in big bang cosmology,  one typically requires about 60 number of e-folds, i.e. $N_f \sim 60$.

%%%%%%%%%%%%%%%%%%%%

To employ $\delta N$ formalism, one has to count the differences in $N$ between two hypersurfaces.   The initial hypersurface is spatially flat where the inflaton perturbations $\delta \phi$ at the moment of horizon crossing have the amplitude $H/2 \pi$. On the other hand,  the final hypersurface is the surface of constant energy density which in simple SR setups is determined by the time of reheating. Between these two hypersurfaces, 
one obtains $N$ as a function of the background field and its velocity, 
$N= N(\phi, \dot \phi)$. Correspondingly, the curvature perturbation $\calR$ on superhorizon scales is related to $\delta N$ via  $\calR= \delta N$ \cite{ Lyth:2004gb,  Abolhasani:2019cqw}.

In general, $N(\phi, \dot \phi)$ can be calculated  non-perturbatively. However, in many situations we are interested in perturbative regimes in which, 
after obtaining  $N(\phi, \dot \phi)$, one expends it perturbatively to calculate 
the correlations functions such as the power spectrum, bispectrum, trispectrum etc. Expanding $N(\phi, \dot \phi)$ in this way, yields 
\ba
\label{deltaN1}
\delta N &=& N' \delta \phi + \frac{N''}{2!} \delta \phi^2 + \frac{N'''}{3!} \delta \phi^3+... \nonumber\\
&=& \sum_{j=1}^\infty N^{(j)}(\phi, \dot \phi)
   \frac{\delta\phi^j}{j!} \, .
\ea
Here and below, a prime on $N$ indicates the derivative with respect to the background $\phi$, i.e.  $N'= N_{, \phi}, N''= N_{, \phi \phi}$ and so on. Similarly,  $N^{(j)}$ represents the $j$-th derivative with respect to $\phi$. 

In writing 
the expansion in Eq. (\ref{deltaN1}) we have neglected $\delta \dot \phi$ because it vanishes on superhorizon scales. Having said this, it is crucial to note that we should treat $N$ as a function of both $\phi$ and $\dot \phi$ so it is a function in the two-dimensional phase space. Only in the case of attractor regime one can interchangeably use $t \leftrightarrow \phi$ and consider $N$ as function of $\phi$. However, in general, as in USR setup when the system is in non-attractor regime, this assumption is not valid and one has to solve $N$ in the full phase space. 

With the relation $\calR = \delta N$ and using the latter expansion given in 
Eq. (\ref{deltaN1}), one can provide a direct link with the correlations of $\calR$ and $\delta \phi$ to all orders in perturbation theory. This can be done either in configuration space or in Fourier space. Usually, it is more convenient to work in the Fourier space ($\bfk$ space) in which the amplitude of $\delta \phi_\bfk$ is known at the moment of horizon crossing,
\ba
\label{phi-powe}
\langle \delta \phi_{\bfk_1}  \delta \phi_{\bfk_2} \rangle =
( 2 \pi)^3  \delta^3(\bfk_1+ \bfk_2)
\frac{H^2}{ 2 k^3}  \equiv ( 2 \pi)^3  \delta^3(\bfk_1+ \bfk_2) P_{\delta \phi} \, .
\ea
In using $\delta N$ formalism it is understood that all modes of interest $\bfk_i$ are superhorizon, $k_i < a H$.

Starting with the power spectrum, 
\ba
\langle \calR_{\bfk_1}  \calR_{\bfk_2} \rangle \equiv ( 2 \pi)^3  \delta^3(\bfk_1+ \bfk_2) P_{\calR}
\ea
we obtain, 
\ba
\label{power-eq}
P_\calR(\bfk) = (N')^2 P_{\delta \phi} = (N')^2 \frac{H^2}{2 k^3}  \, . 
\ea

Going to higher order correlation functions,  the bispectrum is given by the three-point function via,
\ba
\langle \calR_{\bfk_1} \calR_{\bfk_2} \calR_{\bfk_3}\rangle = (2 \pi)^3 \delta^3(\bfk_1+ \bfk_2+ \bfk_3)  B_\calR(k_1, k_2, k_3)  \, .
\ea
As the inflaton perturbations $\delta \phi$ is Gaussian, to calculate the bispectrum one has to consider the second order expansion in  Eq. (\ref{deltaN1}) in which two $\delta \phi$'s are contracted with one  $(\delta \phi)^2$. 
Performing the corresponding contractions associated to Gaussian perturbations $\delta \phi$, one obtains   
\ba
\label{BR-1}
B_\calR(k_1, k_2, k_3) = (N')^2 N''  \Big( \frac{H}{2} \Big)^4 S_{(3)}(k_1, k_2, k_3) \, ,
\ea
in which the bispectrum shape is defined via,
\ba
\label{S3-eq}
S_{(3)}\equiv \frac{1}{k_1^3}  \frac{1}{k_2^3} +  \frac{1}{k_1^3}  \frac{1}{k_3^3} + \frac{1}{k_2^3}  \frac{1}{k_3^3} \, .
\ea
This is the standard local-shape non-Gaussianity. An immediate conclusion 
 is that the bispectrum has only one independent shape (the local shape)

It is customary to define the amplitude of local-type non-Gaussianity $f_{NL}$ by factoring out two powers of $P_\calR$ from Eq. (\ref{BR-1}). Noting from Eq. (\ref{power-eq}) that $P_\calR$ is related to $P_{\delta\phi}$ by an additional factor $N'^2$, the bispectrum is written in the usual form as follows  \cite{Byrnes:2006vq},
\ba
 B_\calR(k_1, k_2, k_3)  = \frac{6}{5} f_{NL} 
 \Big[P_\calR(k_1) P_\calR(k_2)  + P_\calR(k_1) P_\calR(k_3)+ P_\calR(k_2) P_\calR(k_3) \Big] \, ,
\ea
in which $f_{NL}$ is obtained to be, 
\ba
\label{fNL-eq}
f_{NL}= \frac{5}{6} \frac{N'' (N')^2}{(N')^4} = \frac{5}{6} \frac{N'' }{(N')^2}
\equiv \frac{5}{6} F_{NL} \, .
\ea
For the later convenience, we have defined $F_{NL}$ by absorbing the numerical factor $5/6$.

The last example is the trispectrum which is given by the  four-point function as follows,
\ba
\label{trispectrum}
\langle \calR_{\bfk_1} \calR_{\bfk_2} \calR_{\bfk_3} \calR_{\bfk_4} \rangle = (2 \pi)^3 \delta^3(\bfk_1+ \bfk_2+ \bfk_3+ \bfk_4)  T_\calR(k_1, k_2, k_3, k_4) \, .
\ea
To calculate the trispectrum we have two options in higher order contractions. Either contract two $\delta \phi$ with one $\delta \phi^2$, or contact three $\delta \phi$ with one $\delta \phi^3$. Correspondingly, in Fourier space,  we have two distinct local shapes in trispectrum, parameterized as follows  \cite{Byrnes:2006vq}, 
\ba
\label{TR}
T_\calR(k_1, k_2, k_3, k_4) &=& \tau_{NL} \big[ P_\calR(k_{13}) P_\calR(k_3) P_\calR(k_4) + (11~ \text{perms}) \big] \nonumber\\
&+& \frac{54}{25} g_{NL} \big[ P_\calR(k_{2}) P_\calR(k_3) P_\calR(k_4)
+ (3~ \text{perms}) \big] \, ,
\ea
where we have defined  $k_{ij} \equiv | \bfk_i+\bfk_j |$. The two distinct  shapes of trispectrum from the above expression are defined as follows,  
\begin{equation}
%\label{S4-eq}
\label{tri-shapes}
\begin{split}
    &S_{(4)}^{(a)}(\bfk_1,\bfk_2,\bfk_3, \bfk_4)=\left(\frac{1}{k_1^3k_2^3k_3^3}+ \text{3 Permutations }\right) \, ,\\
     &S_{(4) }^{(b)}(\bfk_1, \bfk_2, \bfk_3, \bfk_4)=\left(\frac{1}{k_1^3k_2^3k_{13}^3}+\text{11 Permutations}\right) \, .
     \end{split}
\end{equation}
In Eq. (\ref{TR}),  $\tau_{NL}$ and $g_{NL}$ are two new parameters associated to the  local-shape trispectrum. After taking into account the additional factor $(N')^2$ relating $P_{\delta \phi}$ to $P_\calR$, we obtain  
\ba
\label{gNL}
g_{NL} = \frac{25}{54} \frac{N''' (N')^3}{(N')^6} =\frac{25}{54} \frac{N'''}{(N')^3}
\equiv  \frac{25}{54} G_{NL} \, ,
\ea
and 
\ba
\label{tauNL}
\tau_{NL} = \frac{(N'')^2 (N')^2}{(N')^6}  = \frac{(N'')^2}{(N')^4} \equiv T_{NL} \, .
\ea
Again, to absorb the numerical prefactors and to unify the notation, we have defined $G_{NL}$ and $T_{NL}$. 

Upon inspection, one observes that  $\tau_{NL}$ and $f_{NL}$
in Eqs. (\ref{tauNL}) and (\ref{fNL-eq}) are related to each other via 
$T_{NL}= F_{NL}^2$. This is the special case of the Suyama-Yamaguchi (SY) inequality which states that $T_{NL} \ge  F_{NL}^2$ for general multiple field setups \cite{Suyama:2007bg, Suyama:2010uj} . 
The equality in the above inequality  is valid for any single field model of inflation in the tree level i.e. in the absence of loop corrections  \cite{Tasinato:2012js}. 

Motivated by the above examples, in this work we study the $n$-point correlation function of curvature perturbation at the tree level
for a general single field model, 
\begin{equation}
\label{n-point}
    \left\langle\mathcal{R}_{\bfk_1}\mathcal{R}_{\bfk_2}....\mathcal{R}_{\bfk_n}\right\rangle \,  .
\end{equation}
As the example of trispectrum has shown, one expects to have more independent shapes for higher values of $n$. In addition, one may expect a generalized form of the SY equality to emerge. 
In order to study the higher order correlation functions, in the next section we present a graph theory representation of the general 
$n$-point correlation function. This approach, as in QFT Feynman diagrams, helps us to visualize the important ingredients in higher order correlations and to classify independent shapes based on their physical and mathematical structures.   

Before presenting our graph theory diagrams for $\delta N$ analysis, 
we comment that diagrammatic representations of cosmological correlators and non-Gaussianity were studied in  \cite{Byrnes:2007tm, Tasinato:2012js, Yokoyama:2008by, Suyama:2011qi} as well. 
In particular, Refs. \cite{Byrnes:2007tm, Tasinato:2012js} employ Feynman-type diagrammatic approach to study $n$-point functions with emphasis on loop corrections. On the other hand, Refs. \cite{ Yokoyama:2008by, Suyama:2011qi} study $n$-point functions using diagrammatic approach similar to ours. Ref. \cite{Yokoyama:2008by} presents reduced and tree diagrams with the aim to calculate the contributions 
to the $n$-point correlation functions at a given time $N_F$ in the tree diagrams. The vertices start from an initial flat hypersurfaces and, by use of interacting vertices as well as propagators, they are connected to the time $N_F$. The contribution of  each tree to the $n$-point correlation function is then given as an integral over the production of the propagators and interacting vertices \cite{Yokoyama:2008by}. On the other hand, Ref. \cite{Suyama:2011qi} studies the SY inequality for higher $n$-point correlation functions 
in which one can write inequalities between the amplitudes of 
some certain diagrams and two sub diagrams obtained from them.

%%%%%%%%%%%%%%%%%%%%%%%%%%%%%%%%%
\section{$\delta N$ Diagrams}

Here we present our graph theory representation of $\delta N$ formalism in calculating the $n$-point correlation function. %Eq. (\ref{n-point}).

Looking at the $\delta N$ expansion in terms of $\delta \phi$ in Eq. (\ref{deltaN1}), we see that due to the nonlinear form of the expansion, various powers of $\delta \phi$'s can  contract with each other to yield a nonzero value for the $n$-point correlation function. This is clearly seen in the example of  bispectrum studied in the previous section in which one has to contract two $\delta \phi$'s with external momentums $\bfk_i$ and $\bfk_j$ with a quadratic combination $(\delta \phi^2)_{\bfk_r}$ subject to $\bfk_i+ \bfk_j+ \bfk_r=0$. On the other hand, the situation for trispectrum becomes more non-trivial as we have two different non-linear combinations $\delta \phi^2$ and $\delta \phi^3$ in which the remaining $\delta \phi$'s can contract. Correspondingly, we obtain two different shapes for the trispectrum controlled by the parameters $G_{NL}$ and $T_{NL}$. As one can see, going to higher order this pattern repeats and more an more independent shapes will emerge. To study the general case of correlation functions, it proved very helpful to look at the correlation functions as in graph theory diagrams which we will elaborate below.

%%%%%%%%%%%%%%%%%%%%%%%%%%%%%%%%%%%%%%
\subsection{Definitions and basic graph-theoretic results}
\label{defs-sec}

We begin by introducing the graph-theoretic terminology needed to classify the connected tree level contributions to the \(n\)-point correlation functions.

In our diagrammatic representation, each curvature perturbation \(\mathcal R_{\mathbf k_i}\) appearing in the \(n\)-point function is represented by a vertex. Expanding \(\mathcal R_{\mathbf k_i}\) in powers of the Gaussian field perturbation \(\delta\phi\), as in Eq. (2.2), each vertex is associated with a term proportional to \(N^{(d)}(\delta\phi)^d/d!\). Wick contractions between field perturbations belonging to different vertices are represented by edges connecting the corresponding vertices. Consequently, a vertex associated with the \(d\)-th order term in the \(\delta N\) expansion has \(d\) incident edges. The resulting diagrams can therefore be studied using the standard concepts of graph theory. In the following, we restrict ourselves to connected tree level diagrams. These are represented by graphs with vertices corresponding to curvature perturbation insertions and edges corresponding to Gaussian contractions. The momentum labels are temporarily suppressed when classifying the graph topologies.\\

\begin{definition}
\rm{(Degree and degree sequence)}:
 The degree \(d_v\) of a vertex \(v\) is the number of edges incident to it. The degree sequence of a graph with \(n\) vertices is the sequence \((d_1,\ldots,d_n)\) of vertex degrees arranged in non-increasing order, \(d_1\geq d_2\geq\cdots\geq d_n\). In our \(\delta N\) representation, a vertex of degree \(d\) corresponds to the term \(N^{(d)}(\delta\phi)^d/d!\) in the expansion of the curvature perturbation.
\end{definition}

\begin{definition}
\rm (Path). A path is a sequence of distinct vertices \(v_0,v_1,\ldots,v_r\) such that each pair of consecutive vertices is connected by an edge.
\end{definition}

\begin{definition}
\rm (Connected graph): A graph is connected if there exists a path between every pair of distinct vertices.
\end{definition}

\begin{definition}
\rm (Cycle): A cycle is a closed sequence of edges connecting distinct vertices, except that the initial and final vertices coincide.
\end{definition}

\begin{definition}
\rm (Tree): A tree is a connected graph containing no cycles.
\end{definition}

\begin{definition}
\rm (Leaf):  A leaf is a vertex of degree one. In our \(\delta N\) representation, a leaf corresponds to the linear term \(N'\delta\phi\).
\end{definition}

\begin{definition}
 \rm (Star tree):  A star tree $S_n$ is a tree with $n\geq3$ vertices in which one vertex is adjacent to all the remaining \(n-1\) vertices. Its degree sequence is $(n-1,1,\ldots,1)$. In our follow up  $\delta N$ representation, we refer to the star tree as the ``primary diagram." The significance of this terminology will become clear in Section \ref{amplitudes-sec}, where we study the factorization of tree amplitudes.
\end{definition}

\vspace{0.5cm}

%%%%%%%%%%%%%%%%%%%%%%%%%%%%%%%%%%%%%%
\begin{figure}[t!]
%\vspace{-1cm}
\centering
%\hspace{-1cm}
    %\includegraphics[scale=0.23]{Tree.png}
    \includegraphics[width=\textwidth]{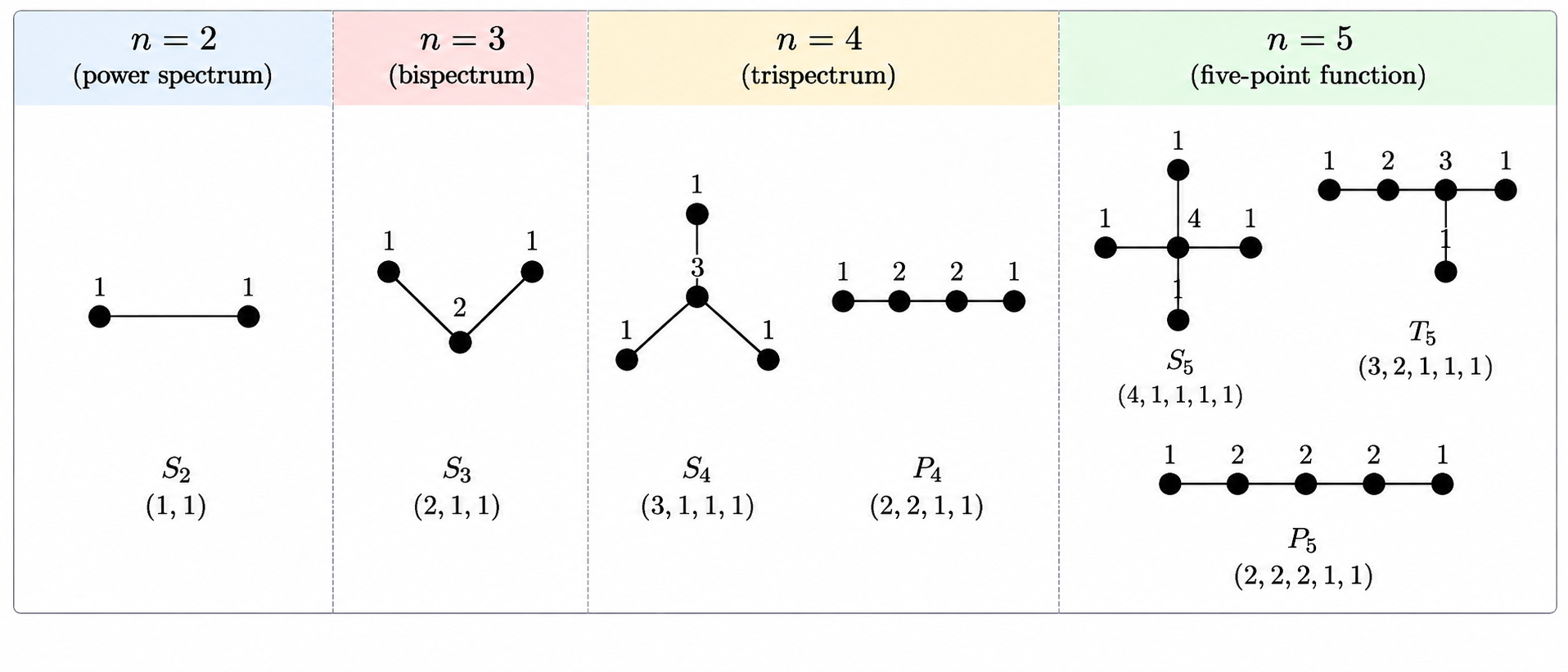}
   % \vspace{0.4cm}
    \caption{Non-isomorphic (independent)  tree diagrams for
    $n=2,3,4,5$. Each vertex represents a curvature
    perturbation insertion, and the number above
    each vertex denotes its degree. The degree
    sequence is displayed beneath each diagram. }
    \label{tree-diags}
\end{figure}

%%%%%%%%%%%%%%%%%%%%%%%%%%%%%%%%%%%%%%
\begin{figure}[h!]
    \centering
   \includegraphics[width=0.85\textwidth]{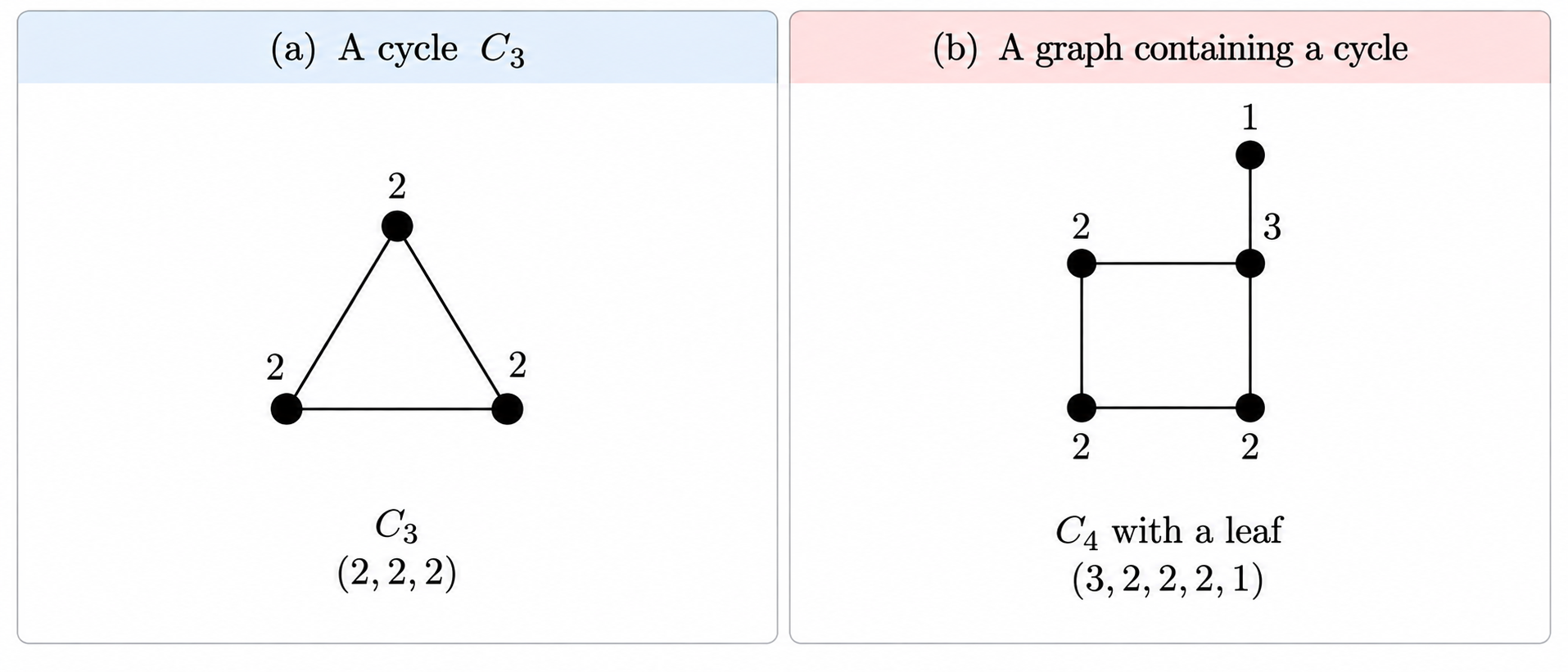}
    \caption{Examples of graphs containing cycles.
    (a) The triangle $C_3$, consisting of three
    vertices of degree two.
    (b) A connected graph consisting of a
    four-vertex cycle $C_4$ with an additional
    leaf attached to one vertex.
    The numbers next to the vertices indicate
    their degrees. 
    }
    \label{cycle}
\end{figure}
%%%%%%%%%%%%%%%%%%%%%%%%%%%%%%%%%%%%%%

It is instructive to examine the simple tree diagrams associated with $n=2,3,4, 5$ before proceeding to the general \(n\)-point correlation functions. These diagrams are illustrated in Fig. \ref{tree-diags}. For the power spectrum (\(n=2\)), the graph consists of two vertices of degree one (leaves), connected by a single edge. For the bispectrum (\(n=3\)), there are three vertices and two edges. The central vertex has degree two, corresponding to the quadratic term \(N''(\delta\phi)^2/2!\) in the \(\delta N\) expansion, while the remaining two vertices are leaves associated with the linear term \(N'\delta\phi\). There is only one  tree diagram with three vertices, corresponding to the unique local bispectrum shape. For the trispectrum (\(n=4\)), there are two independent trees, each containing four vertices and three edges. The first is the star tree \(S_4\), with degree sequence \((3,1,1,1)\), whose central vertex corresponds to the cubic term in the \(\delta N\) expansion. This diagram generates the momentum shape \(S_{(4)}^{(a)}\) in Eq. (\ref{tri-shapes}), with amplitude \(G_{\rm NL}\). The second is the path tree \(P_4\), with degree sequence \((2,2,1,1)\), containing two vertices of degree two and two leaves. It generates the momentum shape \(S_{(4)}^{(b)}\) in Eq. (\ref{tri-shapes}), with amplitude \(T_{\rm NL}\). 
Finally, for $n=5$ we have three independent diagrams identified by the sequences $(4,1,1,1,1)$ which is the star diagram $S_{5}$, $(3,2,1,1,1)$ which is denoted by $T_5$ and $(2,2,2, 1,1)$ which is the path tree $P_5$.

The notation in Fig. \ref{tree-diags} follows standard graph-theoretic terminology: \(S_n\) denotes a star tree with \(n\) vertices, while \(P_n\) denotes a path graph with \(n\) vertices, in which the vertices form a single chain. The symbol \(T_5\) is used in Fig. \ref{tree-diags} as a label for the remaining five-vertex tree, which is neither a star nor a path; it has degree sequence \((3,2,1,1,1)\). All diagrams in Fig. \ref{tree-diags} are connected and contain no cycles, and therefore qualify as trees. This restriction is important for our analysis, since we consider the connected tree level contributions to the \(\delta N\) correlation functions
and  do not study quantum loop corrections in correlation functions.   For comparison, Fig.  \ref{cycle} illustrates graphs containing cycles. In standard graph theory, \(C_n\) denotes a cycle graph with \(n\) vertices, each of degree two. Figure \ref{cycle}(a) shows the triangle \(C_3\), while Fig. \ref{cycle}(b) shows a cycle \(C_4\) with an additional leaf attached to one of its vertices. These graphs are not trees because they contain cycles, and hence lie outside the class of diagrams considered in our tree level analysis.

With the above definitions, we first consider the relation between the number of vertices and edges in a tree. A standard result in graph theory states the following \cite{bondy}.\\

{\bf Theorem 1.} A tree with \(n\) vertices has exactly \(n-1\) edges.

This statement can be verified directly from Fig. \ref{tree-diags}, where each tree with \(n=2,3,4,5\) vertices contains precisely \(n-1\) edges. In our \(\delta N\) representation, each edge corresponds to a contraction of two field perturbations, or equivalently to a propagator of \(\delta\phi\). Thus, a connected tree level $n$-point function contains exactly \(n-1\) 
propagators.\\

The above theorem also implies the following characterization of trees.\\

{\bf Lemma 1.}  Among all connected graphs with \(n\) vertices, the minimum possible number of edges is \(n-1\). This minimum is attained if and only if the graph is a tree.

Indeed, every connected graph contains a spanning tree, which has \(n-1\) edges. If the graph contains additional edges, it necessarily contains a cycle. The graphs in Fig. \ref{cycle} illustrate this distinction in which  both diagrams are connected but contain cycles and therefore have more than \(n-1\) edges.\\

Next, we consider the relation between the number of propagators and the degrees of the vertices. Let \(d_v\) denote the degree of vertex \(v\). Then one has \cite{bondy},  
\ba
\label{sum-dv}
\sum_{v=1}^{n}d_v=2(n-1).
\ea
In our \(\delta N\) representation, the degree \(d_v\) determines the order of the field perturbation at vertex \(v\). Consequently, Eq. (\ref{sum-dv}) shows that the total number of \(\delta\phi\) fields appearing in the corresponding tree level \(n\)-point contribution is \(2(n-1)\), with the fields contracted pairwise to form \(n-1\) propagators.

This relation can be checked explicitly in Fig. \ref{tree-diags}. For the power spectrum, \(n=2\) and the degrees are \((1,1)\), giving \(1+1=2\). For the star diagram \(S_4\), associated with one of the trispectrum shapes, the degrees are \((3,1,1,1)\), giving \(3+1+1+1=6=2(4-1)\). The same relation holds for the path diagram \(P_4\), whose degree sequence is \((2,2,1,1)\).\\

We now ask under what conditions a given sequence of positive integers \((d_1,\ldots,d_n)\) can represent the vertex degrees of a tree diagram. Before addressing this question, we present a useful property of trees:

{\bf Lemma 2.} Every tree with at least two vertices has at least two leaves, namely vertices of degree one \cite{bondy}.

For example, all trees in  Fig. \ref{tree-diags} contain at least two leaves. The path trees \(P_n\) have exactly two leaves, whereas the star trees \(S_n\) have \(n-1\) leaves.\\

Using the preceding results, we obtain the following characterization of the tree degree sequences.\\

{\bf Theorem 2.}  A given sequence of \(n\geq2\) positive integers \((d_1,\ldots,d_n)\) is the degree sequence of at least one tree if and only if,
\ba
\sum_{i=1}^{n}d_i=2(n-1).
\ea

{\bf Proof:}  Necessity follows immediately from Eq. (\ref{sum-dv}). To establish sufficiency, we proceed by induction on \(n\). For \(n=2\), the only possible sequence of positive integers satisfying the condition is \((1,1)\), which corresponds to the single edge tree representing the power spectrum. Now, suppose the statement holds for sequences of \(n\) positive integers. Consider a sequence \((d_1,\ldots,d_{n+1})\) satisfying $\sum_{i=1}^{n+1}d_i=2n$.  As in the proof of Lemma 2, this condition implies that at least two entries are equal to one. Choose one such entry, say \(d_j=1\), and another entry \(d_k\geq2\). Remove \(d_j\) from the sequence and replace \(d_k\) by \(d_k-1\). The resulting sequence contains \(n\) positive integers and satisfies $\sum_{i=1}^{n}d'_i=2(n-1)$. By the induction hypothesis, this reduced sequence is realized by a tree with \(n\) vertices. Attaching a new leaf to the vertex whose degree was reduced from \(d_k\) to \(d_k-1\) restores the original degree sequence without introducing a cycle. The resulting graph is therefore a tree with \(n+1\) vertices and the prescribed degrees. This completes the induction. \(\square\)\\

Theorem 2 establishes that every sequence of positive integers satisfying Eq. (\ref{sum-dv}) can be realized by at least one tree. However, the degree sequence does not necessarily determine the tree uniquely. In particular, for sufficiently large \(n\), distinct tree topologies may share the same degree sequence. This distinction becomes relevant for the classification of 
higher order correlation functions. We therefore introduce the standard graph-theoretic notion of 
isomorphism:\\

{\bf Definition 8:} (Graph isomorphism). Two trees \(T_1=(V_1,E_1)\) and \(T_2=(V_2,E_2)\) are isomorphic if there exists a bijection \(f:V_1\to V_2\) such that, for every pair of vertices \(u,v\in V_1\),
\ba
\{u,v\}\in E_1
\quad\Longleftrightarrow\quad
\{f(u),f(v)\}\in E_2 \, ,
\ea 
in which $\{x,y\}$ represents the edge connecting the vertices $x$ and $y$.
In other words, an isomorphism preserves adjacency between vertices. Two trees are non-isomorphic if no such bijection exists.

An immediate consequence is that isomorphic trees have the same vertex degrees, up to permutation. The converse, however, is not generally true. As an example, consider the two trispectrum diagrams in Fig. \ref{tree-diags}. The star tree \(S_4\) has degree sequence \((3,1,1,1)\), while the path tree \(P_4\) has degree sequence \((2,2,1,1)\). Since graph isomorphisms preserve vertex degrees, these two trees cannot be isomorphic. They therefore represent two distinct tree topologies, associated with the two independent local trispectrum shapes discussed in Section \ref{review}.

This distinction motivates the counting of non-isomorphic trees, which we consider in the next subsection.

\subsection{Number of Independent Diagrams }
\label{shape-number}
%%%%%%%%%%%%
%%%%%%%%%%%%

The first question we address is the number of independent tree diagrams for a given \(n\), where independent diagrams are understood as non-isomorphic trees. Although there is no simple closed form expression for this number, the following generating function relations, associated with the Otter theorem \cite{otter}, allow us to determine it at arbitrary order.\\

{\bf Theorem 3} (Otter's tree counting theorem). Let \(T_{(n)}\) denote the number of non-isomorphic trees with \(n\) vertices. Define the generating function via, 
\ba
\label{phi(x)}
\phi(x)=T_{(1)}+T_{(2)}x+T_{(3)}x^2+\cdots.
\ea
Then,
\ba
\label{phi(x)2}
{\phi(x)=\psi(x)-\frac{x}{2}\left[\psi(x)^2-\psi(x^2)\right],}
\ea
where
\ba
\psi(x)=A_1+A_2x+A_3x^2+\cdots \, ,
\ea
is the corresponding generating function, with \(A_n\) denoting the number of such trees with \(n\) vertices. It satisfies
\ba
\label{psi(x)}
{\psi(x)=\exp\Big[\sum_{v=1}^{\infty}\frac{x^v}{v}\psi(x^v)\Big].}
\ea
Solving the above recurrence relations together yields the number of non-isomorphic trees associated with $n$.

To illustrate this procedure, let us expand \(\psi(x)\) to the first few orders,
\ba
\psi(x)=A_1+A_2x+A_3x^2+O(x^3).
\ea
On the other hand, expanding Eq. (\ref{psi(x)}), we obtain,
\ba
\begin{aligned}
\psi(x)
=1+A_1x
+\frac{A_1^2+A_1+2A_2}{2}x^2
+O(x^3).
\end{aligned}
\ea
Matching the coefficients of equal powers of \(x\) in the above two equations yields,
\ba
A_1=1,\qquad A_2=1,\qquad A_3=2.
\ea
Substituting these results into Eq. (\ref{phi(x)2}), we find,
\ba
\phi(x)=1+x+x^2+O(x^3),
\ea
which gives \(T_{(1)}=T_{(2)}=T_{(3)}=1\). Therefore, there is one independent tree diagram for the power spectrum (\(n=2\)) and one tree diagram for the bispectrum (\(n=3\)) as illustrated in Fig. \ref{tree-diags}.  Here, \(T_{(1)}=1\) counts the trivial tree consisting of a single vertex.

Continuing this procedure to higher orders, we obtain
\ba
T_{(4)}=2,\qquad T_{(5)}=3,\qquad T_{(6)}=6, \qquad
T_{(7)}=11,\qquad T_{(8)}=23,\quad T_{(9)}=47,\quad
T_{(10)}=106.
\ea
The two independent diagrams for the trispectrum (\(T_{(4)}=2\)) and the three independent diagrams for the five-point function (\(T_{(5)}=3\)) are displayed in Fig. \ref{tree-diags}. 

As can be seen from the above examples, the number of independent diagrams increases rapidly with \(n\). For large \(n\), Otter's asymptotic result gives \cite{otter}, 
\ba
\label{Otter-large-n}
T_{(n)}\sim
{0.535\,(2.956)^n} \, {n^{-\frac{5}{2}}},
\qquad \qquad  (n\gg1) \, .
\ea
For example, at \(n=20\), the exact number of non-isomorphic trees is \(T_{(20)}=823\,065\), whereas the asymptotic expression gives approximately \(776\,000\), differing from the exact result by about \(5.7\%\).

%We emphasize that \(T_{(n)}\) counts the number of distinct tree topologies, independently of vertex labeling. Their correspondence with distinct momentum shapes is further discussed in Section \ref{momentum-shapes}. Although the number of tree topologies grows rapidly with \(n\), their amplitudes have a considerably simpler structure, as we demonstrate in next Section.  

%%%%%%%%%%%%
%%%%%%%%%%%%

\section{ Tree Amplitudes and Their Factorization}
\label{amplitudes-sec}

In this section, we study the amplitudes associated with the connected tree diagrams contributing to the $n$-point correlation functions. As in the examples of the bispectrum and trispectrum discussed in Section \ref{review}, the contribution of each tree diagram can be separated into an amplitude and a momentum dependent shape function. Our main goal here is to establish the general structure of the amplitudes and their relations at
different orders. %The explicit construction of the Fourier-space shape functions will be presented in \ref{momentum-shapes}.

%%%%%%%%%%%%%%%%%%%%%%%%%%%%%%%%%%%%
\subsection{Tree Amplitudes}

Consider a tree $T$ with $n$ vertices, and denote the degree of vertex $v$ by $d_v$. As established in Section \ref{defs-sec}, the degrees satisfy,
\begin{equation}
\sum_{v\in V(T)}d_v=2(n-1).
\label{degree-sum}
\end{equation}
The contribution of this tree to the connected $n$-point
function is proportional to
\begin{equation}
\left\langle
{\cal R}_{\mathbf{k}_1}\cdots {\cal R}_{\mathbf{k}_n}
\right\rangle_T' = \Big(\frac{H}{2}\Big)^{2(n-1)}
\prod_{v\in V(T)}N^{(d_v)} \, S_T(\bfk_1,..., \bfk_n) \, ,
\label{tree-S-amp}
\end{equation}
where the prime indicates that the overall momentum conserving delta function has been omitted. Here, $S_T(\bfk_1,..., \bfk_n)$ represents the shape function which will be studied in 
Section \ref{momentum-shapes}. Our goal in this section is to concentrate on the amplitude of the correlation function.

As in the cases of bispectrum and trispectrum, we can define the amplitude of the $n$-point function associated with the tree $T$ with the degree 
sequence $(d_1,...,d_n)$. Noting that $P_\calR$ scales like $N'^2 P_{\delta\phi}$, from Eq. (\ref{tree-S-amp}) we define the amplitude 
$f_T$ associated to that shape function as follows,
\ba
f_T\equiv
\frac{\displaystyle\prod_{v\in V(T)}N^{(d_v)}}
{(N')^{2(n-1)}}.
\label{fT-eq}
\ea
In addition, it is convenient to introduce,
\ba
a_d\equiv\frac{N^{(d)}}{(N')^d},
\qquad \qquad a_1=1.
\label{ad-def}
\ea
We refer to \(a_d\) as the normalized vertex amplitude associated with a vertex of degree \(d\). The normalization ensures that \(a_1=1\), so that leaves contribute a unit factor to the tree amplitude. 

Using the graph-theoretic degree sum relation Eq. (\ref{degree-sum}), the tree amplitude takes the factorized form,
\ba
f_T=\prod_{v\in V(T)}a_{d_v}.
\label{fT-ad}
\ea
This is a fundamental result of our graph-theoretic formulation: the amplitude of any connected tree diagram factorizes into the product of the vertex amplitudes associated with its vertices.

For a tree with degree sequence $(d_1,\ldots,d_n)$, we may equivalently denote its amplitude $f_T$ by $ f_{(d_1,\ldots,d_n)}$. 
For example, comparing with the bispectrum and trispectrum amplitudes introduced in Section \ref{review}, we have in particular, 
\ba
f_{(2,1,1)}=F_{NL}=a_2,
\ea
and
\ba
f_{(3,1,1,1)}=G_{NL}=a_3,
\qquad
f_{(2,2,1,1)}=T_{NL}=a_2^2.
\ea

An important consequence of Eq.~\eqref{fT-ad}
is that the tree amplitude depends only on the degree sequence of the tree. Therefore, two non-isomorphic trees with the same degree sequence have identical  amplitudes, even though their momentum dependent shape functions may differ. %This distinction becomes important for higher order correlation functions and will be discussed in Section \ref{momentum-shapes}. 

%%%%%%%%%%%%%%%%%%%%%%%%%%%%%%%%
\subsection{Leaf gluing and amplitude factorization}
We now introduce an operation that relates the amplitudes
of trees at different orders. Consider two disjoint trees
$T_1$ and $T_2$, each containing at least three vertices.
Choose one leaf in each tree and remove the selected
leaves together with their incident edges. We then
connect their former neighbouring vertices by a new edge.
We refer to this operation as \emph{leaf gluing}. The resulting graph is again a tree. Furthermore,
the degrees of all surviving vertices remain unchanged:
each of the two neighbouring vertices loses one edge
and gains one new edge. This leads to the following theorem.\\

{\bf Theorem 4}: (Leaf-gluing factorization).
Let $T_1$ and $T_2$ be two trees with $m\geq3$ and $n\geq3$ vertices respectively. Let $T_3$ be the tree obtained by leaf gluing $T_1$ and $T_2$. Then $T_3$ contains $m+n-2$ vertices, and its  amplitude satisfies,
\ba
f_{T_3}=f_{T_1}f_{T_2}.
\label{leaf-gluing-eq}
\ea

\noindent
{\bf Proof:}
By construction, the two removed leaves have degree one and hence contribute factors $a_1=1$ to the amplitudes of the original trees. All surviving vertices preserve their degrees. Consequently, the product of vertex factors in $T_3$ is precisely the product of the vertex factors in $T_1$ and $T_2$. Applying Eq.~\eqref{fT-ad},
we obtain
\ba
f_{T_3}
=\prod_{v\in V(T_3)}a_{d_v}
=\Big(\prod_{v\in V(T_1)}a_{d_v}\Big)
 \Big(\prod_{v\in V(T_2)}a_{d_v}\Big)
=f_{T_1}f_{T_2}.
\ea
This completes the proof. \hfill $\square$

\medskip

The above theorem establishes an exact factorization relation
among  tree amplitudes at different orders. The simplest example reproduces the familiar single field SY equality.

%%%%%%%%%%%%%%%%%%%%%%%%%%%%%%%%%%%%%%
\begin{figure}[t!]
%\vspace{-1cm}
\centering
%\hspace{-1cm}
    %\includegraphics[scale=0.23]{Tree.png}
    \includegraphics[width=\textwidth]{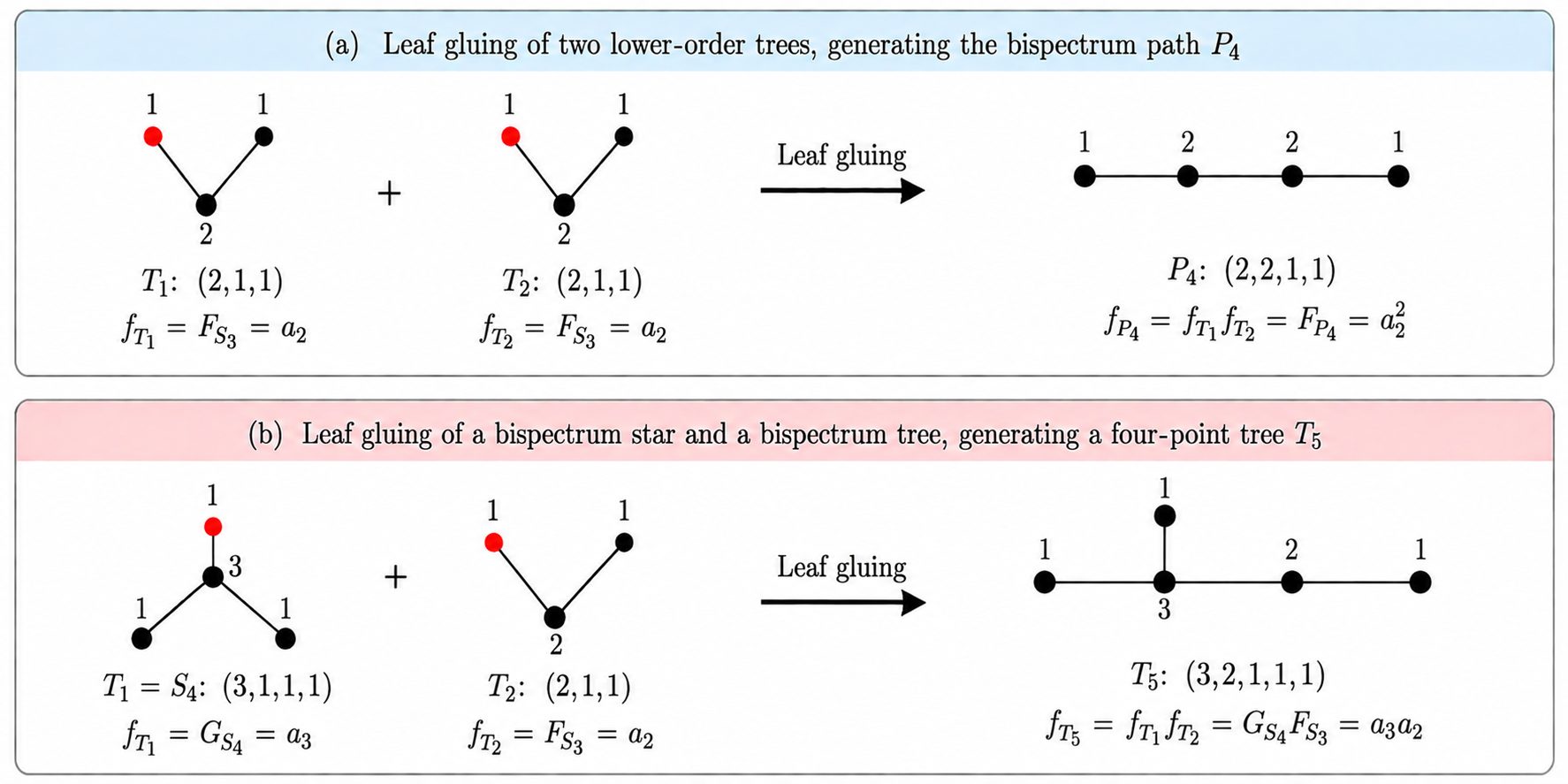}
    \vspace{0.4cm}
  \caption{Examples of the leaf-gluing operation and the
factorization of  tree amplitudes.
The red vertices denote the leaves selected for removal.
In each case, the two selected leaves and their incident edges
are removed, and their former neighbouring vertices are connected
by a new edge. The degrees of all surviving vertices are
preserved. (a) Gluing two bispectrum trees with degree sequence
$(2,1,1)$ produces the four-point path tree $P_4$,
with degree sequence $(2,2,1,1)$. The corresponding
amplitude satisfies $f_{P_4}=F_{NL}^2=a_2^2=T_{NL}$,
reproducing the SY equality. 
(b) Gluing the four-point star $S_4$, with degree sequence
$(3,1,1,1)$, and a bispectrum tree produces the
five-point tree with degree sequence $(3,2,1,1,1)$.
Its  amplitude satisfies
$f_{T_5}=G_{NL}F_{NL}=a_3a_2$.
}
\label{fig:leaf-gluing}
\end{figure}
%%%%%%%%%%%%%%%%%%%%%%%%%%%%%%%%%%%%

In Fig.~\ref{fig:leaf-gluing}, we have illustrated the leaf-gluing operation with two examples. In both cases, the red vertices represent the leaves selected for removal. The operation preserves the degrees of all remaining vertices, and consequently the  amplitude of the resulting tree is the product of the original amplitudes. In the first example, for the four-point function, the path tree $P_4$ is obtained by gluing two bispectrum trees, each with degree sequence $(2,1,1)$. Applying the above theorem yields,
\begin{equation}
T_{NL}=f_{(2,2,1,1)}
=f_{(2,1,1)}^2
=F_{NL}^2=a_2^2.
\end{equation}
This reproduces the familiar SY equality in our normalization.

As the second example, consider the five-point tree with degree sequence $(3,2,1,1,1)$, shown in the lower panel of Fig.~\ref{fig:leaf-gluing}. It can be constructed
by gluing the four-point star $S_4$ and a bispectrum tree. Its amplitude therefore satisfies,
\begin{equation}
f_{(3,2,1,1,1)}
=f_{(3,1,1,1)}f_{(2,1,1)}
=a_2a_3=G_{NL}F_{NL}.
\end{equation}
%The interesting conclusion is that  the amplitude of this five-point diagram is completely determined by the lower order amplitudes $F_{NL}$ and $G_{NL}$.

These examples demonstrate how the amplitudes of higher order tree diagrams can be expressed in terms of lower order amplitudes. Theorem~4 extends this factorization structure to arbitrary tree order. %We emphasize that this theorem establishes an algebraic identity among tree amplitudes. Its relation to the full momentum-dependent correlation functions requires the corresponding shape functions, which will be studied separately.

%%%%%%%%%%%%%%%%%%%%%%%%%%%%%%%%%%%%
\subsection{Primary diagrams.}

The leaf gluing operation allows us to construct all non-star trees from lower order trees. However, the star diagram is the unique exception. To see this, consider a tree $T$ that is not a star. Such a tree contains at least one edge connecting two vertices of degree greater than one. Removing this edge separates the tree into two connected components. By attaching a new leaf to each of the two exposed endpoints, we obtain two smaller trees, each containing at least three vertices. The original tree can then be reconstructed by leaf gluing these two components. In contrast, every edge of a star is incident to a leaf. Therefore, the star cannot be decomposed into two smaller trees by reversing a nontrivial leaf gluing operation.

We refer to the star tree $S_n$ as the \emph{primary diagram}. It has degree sequence $(n-1,1,\ldots,1)$, and its amplitude is,
\ba
f_{S_n}=f_{(n-1,1,\ldots,1)}
=\frac{N^{(n-1)}}{(N')^{n-1}}
=a_{n-1}.
\ea
In comparison, the path tree \(P_n\), containing \(n-2\) vertices of degree two and two leaves, has the amplitude \(f_{P_n}=a_2^{n-2}\).

Since every non-primary tree can be obtained by repeated leaf gluing, 
Eq. (\ref{leaf-gluing-eq}) implies that its amplitude can be expressed as a product of primary amplitudes. More explicitly, let $m_d(T)$ denotes the number of degree-$d$ vertices in a tree $T$. Then, Eq.~\eqref{fT-ad} gives,
\ba
f_T
=\prod_{d=2}^{n-1}a_d^{m_d(T)}
=\prod_{d=2}^{n-1}
\left(f_{S_{d+1}}\right)^{m_d(T)}.
\label{primary-factorization}
\ea
The degree-one vertices do not contribute because
$a_1=1$.

Equation \eqref{primary-factorization} is the central factorization result of this section. It shows that the amplitudes of all connected tree diagrams are determined by the amplitudes of the primary diagrams. 
The corresponding results for \(3\leq n\leq 7\) are summarized in Table \ref{prim-amp-table}. Notice that different tree topologies can have the same degree sequence and hence the same amplitude. However, the factorization theorem does not identify their momentum dependent shape functions, which generally depend on the full topology.

%%%%%%%%%%%%%%%%%%%%%%%%%%%%%%%%%%%%

\begin{table}[t]
\centering
\renewcommand{\arraystretch}{1.25}
\setlength{\tabcolsep}{5pt}
\small
\begin{tabular}{|c| |l | l|}
\hline
$n$ & tree amplitudes
    & Products of primary amplitudes \\
\hline
$3$ &
$~~ f_{(2,1,1)}$ &
$~~a_2$ \\
%\addlinespace[7pt]
\hline
$4$ &
$\begin{array}{l}
f_{(3,1,1,1)},\quad f_{(2,2,1,1)}
\end{array}$ &
$\begin{array}{l}
a_3,\quad a_2^2
\end{array}$ \\
%\addlinespace[7pt]
\hline

$5$ &
$\begin{array}{l}
f_{(4,1,1,1,1)},\quad f_{(3,2,1,1,1)},\\
f_{(2,2,2,1,1)}
\end{array}$ &
$\begin{array}{l}
a_4,\quad a_3a_2,\\
a_2^3
\end{array}$ \\
%\addlinespace[7pt]
\hline

$6$ &
$\begin{array}{l}
f_{(5,1,1,1,1,1)},\quad f_{(4,2,1,1,1,1)},\\
f_{(3,3,1,1,1,1)},\quad f_{(3,2,2,1,1,1)},\\
f_{(2,2,2,2,1,1)}
\end{array}$ &
$\begin{array}{l}
a_5,\quad a_4a_2,\\
a_3^2,\quad a_3a_2^2,\\
a_2^4
\end{array}$ \\
%\addlinespace[7pt]
\hline

$7$ &
$\begin{array}{l}
f_{(6,1,1,1,1,1,1)},\quad f_{(5,2,1,1,1,1,1)},\\
f_{(4,3,1,1,1,1,1)},\quad f_{(4,2,2,1,1,1,1)},\\
f_{(3,3,2,1,1,1,1)},\quad f_{(3,2,2,2,1,1,1)},\\
f_{(2,2,2,2,2,1,1)}
\end{array}$ &
$\begin{array}{l}
a_6,\quad a_5a_2,\\
a_4a_3,\quad a_4a_2^2,\\
a_3^2a_2,\quad a_3a_2^3,\\
a_2^5
\end{array}$ \\
\hline
\end{tabular}
\caption{Factorization of  tree amplitudes
into primary amplitudes for $3\le n\le 7$.
Each degree sequence is listed once. Distinct tree
topologies may have the same degree sequence and
hence the same  amplitude.}
%\label{tab:amplitudes}
\label{prim-amp-table}
\vspace{0.5cm}
\end{table}

%%%%%%%%%%%%%%%%%%%%%%%%%%%%%%%%%%%%

\subsection{Counting distinct tree amplitudes}
\label{sec:amplitude-counting}

Having established the factorization of tree amplitudes into primary amplitudes, we now compare the number of distinct amplitude products with the number of independent tree topologies. 

Let \(N_{\rm amp}(n)\) denote the number of algebraically distinct products of the primary amplitudes \(a_d\) at order \(n\). Intuitively speaking, \(N_{\rm amp}(n)\) counts the number of distinct products of the primary amplitudes \(a_d\) that can be formed at a given order \(n\), subject to the degree-sum constraint. This should not be confused with the number of available nontrivial primary amplitudes \(a_d\), which is simply \(n-2\). 
According to Eq. (\ref{primary-factorization}), each tree amplitude is specified by the nonnegative integers \(m_d(T)\), which count the number of vertices of degree \(d\). The degree-sum relation 
Eq. (\ref{sum-dv}) then implies,
\ba
\label{N-prod}
\sum_{d=2}^{n-1}(d-1)m_d=n-2 \, .
\ea
Thus, each primary amplitude \(a_d\) contributes a weight \(d-1\), and the distinct amplitude products \(N_{\rm amp}(n)\) correspond to the nonnegative integer solutions of this constraint. These solutions are in one-to-one correspondence with the unrestricted integer partitions of \(n-2\). As an example, consider $n=6$. Then Eq. (\ref{N-prod}) gives 
$m_2+ 2 m_3+ 3 m_4+ 4 m_5=4$. Solving the possible integer solutions yields $(m_5=1, m_i=0); (m_4=1, m_2=1, m_i=0); (m_3=2, m_i=0); (m_3=1, m_2=2,  m_i=0); (m_2=4, m_i=0)$ so in total there are five distinct solutions and therefore $N_{\rm amp}(6)=5$.

Since every admissible degree sequence is realized by at least one tree, as established in Theorem 2, we therefore obtain,
\ba
N_{\rm amp}(n)=p(n-2) \, ,
\ea
where \(p(m)\) is the integer partition function, denoting the number of unrestricted integer partitions of $m$. The results for $3 \leq n \leq 10$
are summarized in Table~\ref{tab:amplitude-counting}. The distinction first appears at $n=6$, where six non-isomorphic trees give only five distinct amplitude products as we demonstrated above. This difference increases rapidly with $n$.

%%%%%%%%%%%%%%%%%%%%%%%%%%%%%%%%%
\begin{table}[t]
\centering
\renewcommand{\arraystretch}{1.12}
\setlength{\tabcolsep}{9pt}
\begin{tabular}{|c ||c |c |c|}
\hline
$n$ & $T_{(n)}$ & $N_{\rm amp}(n)$
& $T_{(n)}-N_{\rm amp}(n)$ \\
\hline
3  & 1   & 1  & 0  \\
4  & 2   & 2  & 0  \\
5  & 3   & 3  & 0  \\
6  & 6   & 5  & 1  \\
7  & 11  & 7  & 4  \\
8  & 23  & 11 & 12 \\
9  & 47  & 15 & 32 \\
10 & 106 & 22 & 84 \\
\hline
\end{tabular}
\caption{Comparison between the number $T_{(n)}$ of
non-isomorphic trees and the number $N_{\rm amp}(n)$
of distinct primary amplitude products.
The difference in the fourth column counts the excess of tree topologies over the number of distinct primary amplitude products.}
\label{tab:amplitude-counting}
\end{table}
%%%%%%%%%%%%%%%%%%%%%%%%%%%%%%%%%

The large-$n$ behaviour of  $N_{\rm amp}(n)$ follows from the
Hardy--Ramanujan asymptotic formula
\cite{HardyRamanujan},
\begin{equation}
N_{\rm amp}(n)\sim
\frac{1}{4\sqrt{3}(n-2)}
\exp\Big[\pi\sqrt{\frac{2(n-2)}{3}}\Big].
\end{equation}
Comparing this expression with Otter's asymptotic formula (\ref{Otter-large-n}), we obtain
\begin{equation}
    \lim_{n\to\infty}
    \frac{N_{\rm amp}(n)}{T_{(n)}}=0.
\end{equation}
Thus, although the number of independent momentum shapes 
$T_{(n)}$ grows exponentially with $n$, the number of distinct amplitude products grows considerably more slowly. The important conclusion is that at high orders, many different tree
topologies therefore share the same  amplitude. \\

%%%%%%%%%%%%%%%%%%%%
%%%%%%%%%%%%%%%%%%%%

%Before ending this section, let us consider the special case where $F_{NL}$ is a constant, i.e. independent of the initial field value and the momentum in the phase space. This is not generic as there are models, such as the USR model with transition to SR stage, that this assumption is violated. Starting with the definition $N^{(2)} = F_{NL} N'^2$, taking the derivative repeatedly, and noting that $F_{NL}$ is a constant in phase space, we obtain
%\ba N^{(k)} =(k-1)!F_{NL}^{k-1}(N')^k, \qquad k\geq2. \ea
%From this equation we conclude that,
%\ba a_d=(d-1)!F_{NL}^{d-1}. \ea
%Substituting the above value of $d_d$ into Eq.~\eqref{fT-ad} and using the relation, $\sum_{v\in V(T)}(d_v-1)=n-2$,  we find that, 
%\ba f_T=F_{NL}^{n-2} \prod_{v\in V(T)}(d_v-1)!, \qquad   \qquad (F_{NL}=\mathrm{constant}) .
%\label{constant-FNL}
%\ea
%Thus, in this special case, all tree amplitudes at a fixed order are proportional to the same power of $F_{NL}$, with their relative coefficients determined entirely by the vertex degrees. 

This completes our discussion of the tree amplitudes  and their factorization. In the next section, we turn to the momentum dependent shape functions
associated with individual tree topologies.

%%%%%%%%%%%%
%%%%%%%%%%%%

\section{Momentum Shapes in Fourier Space}
\label{momentum-shapes}

In the previous section we have shown that the 
amplitude of a tree depends only on its vertex degrees.
Now, as defined in Eq. (\ref{tree-S-amp}), we consider the momentum dependent shape functions, which retain information about the full connectivity of the tree. This distinction first becomes particularly
important at $n=6$ where $T_{(n)} \neq N_{\rm amp}(n)$ so non-isomorphic trees can have the same degree sequence.

Consider a tree $T$ with $n$ vertices carrying external
momenta $\mathbf{k}_1,\ldots,\mathbf{k}_n$, with
$\sum_{i=1}^n\mathbf{k}_i=0$.
Removing an edge $e$, partitions its vertices into two
connected sets $A_e$ and $A_e^c$. The momentum magnitude
carried by that edge is,
\ba
q_e=
\Big|\sum_{i\in A_e}\mathbf{k}_i\Big|
= \Big|\sum_{i\in A_e^c}\mathbf{k}_i\Big|.
\ea
In particular, an edge incident to a leaf labeled $i$
carries $q_e=k_i$.

For a fixed assignment of external labels, the momentum
dependence is the product of the $n-1$ edge propagators.
Summing over distinct assignments gives the shape function as follows, 
\ba
S_T(\mathbf{k}_1,\ldots,\mathbf{k}_n)
=
\sum_{\text{inequivalent labelings}} \, ~
\prod_{e\in E(T)}\frac{1}{q_e^3}.
\ea
Here \(E(T)\) denotes the edge set of the tree \(T\), containing \(n-1\) edges.  Labelings which are related by an automorphism that preserves the graph connectivity are equivalent and are counted only once. Consequently, the number of inequivalent labelings in a tree diagram is  \(n!/|\operatorname{Aut}(T)|\). As a simple example, consider the five-point star \(S_5\), whose central vertex has degree four and whose four leaves have degree one. Its edge set is \(E(S_5)=\{(1,2),(1,3),(1,4),(1,5)\}\), for a representative labeling with vertex 1 at the center. Since the four leaves can be permuted arbitrarily, the automorphism group has order \(|\operatorname{Aut}(S_5)|=4!=24\). Therefore, there are only \(5!/4!=5\) inequivalent assignments of the external momenta, corresponding to the five possible choices of the central vertex. The resulting shape function is given in Eq. (\ref{five-star}) below, while its  amplitude is \(f_{S_5}=a_4\).

In contrast, the  amplitude $f_T$ is given by Eq. (\ref{fT-ad}) as established in Section \ref{amplitudes-sec}. The complete tree contribution carries the corresponding factor $f_T S_T$.

The associated shapes for the bispectrum and trispectrum were given in Section \ref{review}, here we concentrate on the shape functions for 
$n=5$ and $n=6$.

%%%%%%%%%%%%%%%%%%%%%%%%%%%%%%%%%%
\subsection{Five-point functions}
\label{subsec:shape-five}

As we have seen previously, there are three non-isomorphic trees with five
vertices: the star $S_5$, the path $P_5$, and the branched tree with degree sequence $(3,2,1,1,1)$. These are the three graphs already displayed in
the last column in Fig. \ref{tree-diags}.

For compactness, let us define $k_{i_1\cdots i_r}
\equiv \left| \mathbf{k}_{i_1}+\cdots+\mathbf{k}_{i_r}
\right|.$
%\begin{equation} k_{i_1\cdots i_r} \equiv \left| \mathbf{k}_{i_1}+\cdots+\mathbf{k}_{i_r} \right|. \end{equation}
The following expressions give one representative labeling for 
the total \(n!/|\operatorname{Aut}(T)|\) labeling of each topology, followed by the remaining distinct permutations of the external labels.

For the star with center $1$, the edge momenta are $k_2,k_3,k_4,k_5$, so
\begin{equation}
\label{five-star}
S_5^{(a)} = \left(
\frac{1}{k_2^3k_3^3k_4^3k_5^3}
+4\ \text{permutations}
\right),
\qquad
~~~ f_5^{(a)}=a_4.
\end{equation}
For the path $1-2-3-4-5$, the edge momenta are $k_1,k_{12},k_{45},k_5$, giving,
\begin{equation}
\label{eq:shape-five-path}
S_5^{(b)}
=
\left(
\frac{1}{k_1^3k_{12}^3k_{45}^3k_5^3}
+59\ \text{permutations}
\right),
\qquad
f_5^{(b)}=a_2^3.
\end{equation}
Finally, for the edges $(1,2),(1,3),(1,4),(4,5)$, the edge momenta are $k_2,k_3,k_5,k_{45}$, and hence,
\begin{equation}
\label{eq:shape-five-branch}
S_5^{(c)} =
\left(
\frac{1}{k_2^3k_3^3k_5^3k_{45}^3}
+59\ \text{permutations}
\right),
\qquad
f_5^{(c)}=a_3a_2.
\end{equation}
The three topologies therefore give three distinct momentum shapes with their amplitudes  in each case given above.

\subsection{Six-point functions}
\label{subsec:shape-six}

\begin{figure}[t]
\centering
\includegraphics[width=0.9\textwidth]{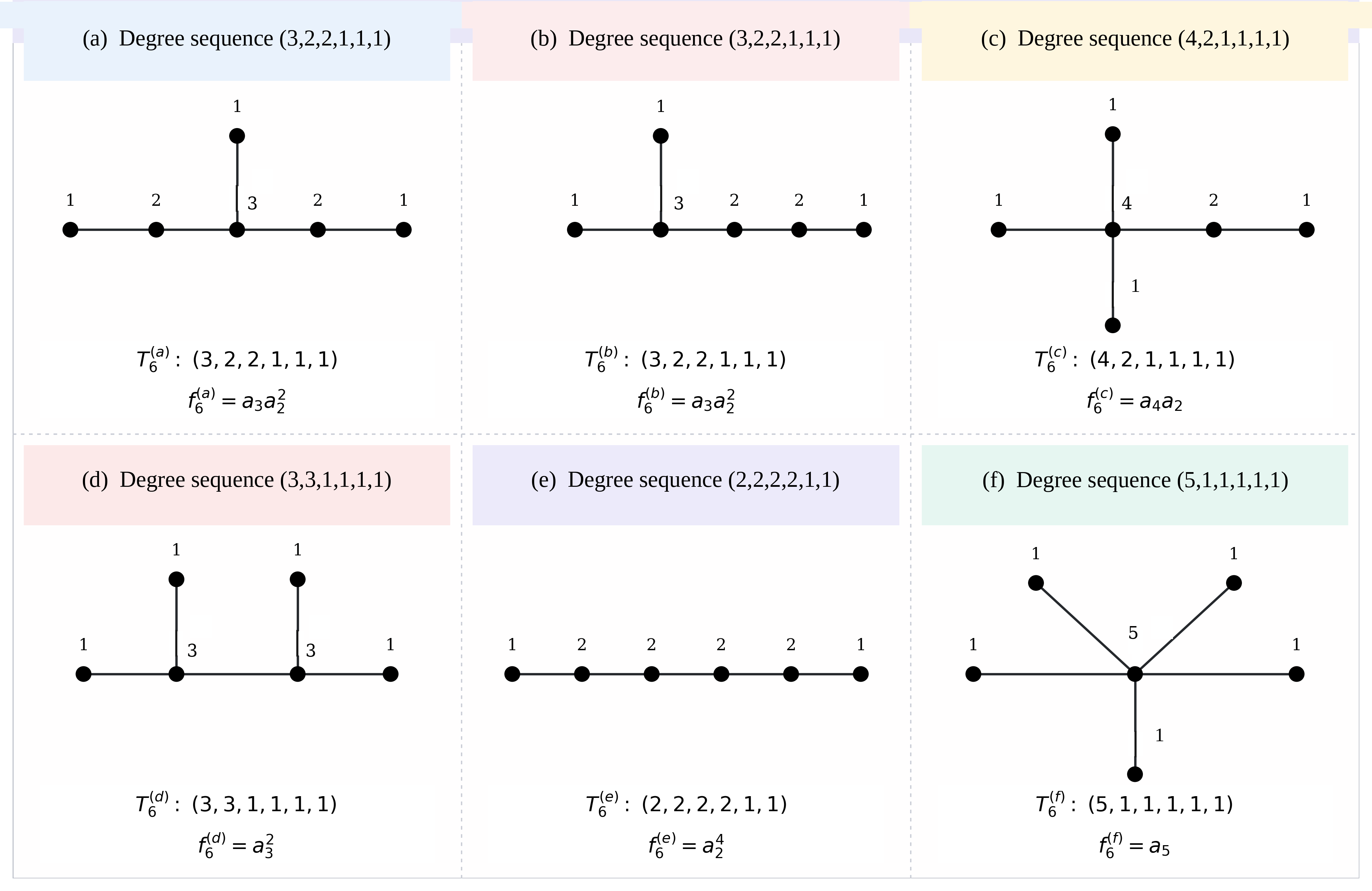}
\caption{
The six non-isomorphic trees associated to $n=6$.  Panels $(a)$ and $(b)$ have identical degree
sequences $(3,2,2,1,1,1)$ and identical tree
amplitudes, but different momentum shapes.
The remaining panels are respectively the
degree $(4,2,1,1,1,1)$ tree, the double-star
$(3,3,1,1,1,1)$, the path $(2,2,2,2,1,1)$,
and the star $(5,1,1,1,1,1)$. 
}
\label{fig:six-point-trees}
\end{figure}

For $n=6$ there are six non-isomorphic trees $N_{(6)}=6$, but only five distinct degree sequences, $N_{\rm {amp}}(6)=5$.  Figure~\ref{fig:six-point-trees} displays all six trees with a representative assignment of
the external labels. The panel letters $(a)$--$(f)$ agree with the shape functions given below. In particular, panels $(a)$ and $(b)$ have the
same degree sequence $(3,2,2,1,1,1)$ but different
edge connectivity.

Reading the edge partitions directly from the
figure, we obtain
\begin{align}
S_6^{(a)}
&=
\left(
\frac{1}{
k_1^3k_5^3k_6^3k_{45}^3k_{126}^3}
+359\ \text{permutations}
\right), \quad \quad ~~~ f_6^{(a)}= a_3 a_2^2
\label{eq:shape-six-a}
\\[2mm]
S_6^{(b)}
&=
\left(
\frac{1}{
k_1^3k_5^3k_6^3k_{12}^3k_{45}^3}
+359\ \text{permutations}
\right), \qquad  ~~~ f_6^{(b)}= a_3 a_2^2
\label{eq:shape-six-b}
\\[2mm]
S_6^{(c)}
&=
\left(
\frac{1}{
k_1^3k_4^3k_5^3k_6^3k_{12}^3}
+119\ \text{permutations}
\right), \qquad ~~~~~ f_6^{(c)}=a_4a_2
\label{eq:shape-six-c}
\\[2mm]
S_6^{(d)}
&=
\left(
\frac{1}{
k_1^3k_4^3k_5^3k_6^3k_{126}^3}
+89\ \text{permutations}
\right), \qquad  ~~~~~ f_6^{(d)}=a_3^2
\label{eq:shape-six-d}
\\[2mm]
S_6^{(e)}
&=
\left(
\frac{1}{
k_5^3k_6^3k_{16}^3k_{126}^3k_{45}^3}
+359\ \text{permutations}
\right), \qquad ~ f_6^{(e)}=a_2^4
\label{eq:shape-six-e}
\\[2mm]
S_6^{(f)}
&=
\left(
\frac{1}{
k_1^3k_2^3k_3^3k_4^3k_5^3}
+5\ \text{permutations}
\right)\, , \qquad ~~~~~~~~~ f_6^{(f)}=a_5 \, .
\label{eq:shape-six-f}
\end{align}

%%%%%%%%%%%%%%%%%%%%%%%%%%%%%%%%%%%%%
%%%%%%%%%%%%%%%%%%%%%%%%%%%%%%%%%%%%%

The equality $f_6^{(a)}=f_6^{(b)}$ illustrates the difference between the amplitude and the shape. The two graphs have identical vertex degrees, but their internal edges produce different momentum partitions. Consequently, their momentum dependence is distinct even though
their amplitudes are equal. More generally, the classification by tree topology determines the distinct momentum structures, while the factorization rule of Section \ref{amplitudes-sec} determines their  amplitudes.

%%%%%%%%%%%%%%%%%
%%%%%%%%%%%%%%%%%

\section{Examples of Single-Field Inflation}
\label{sec:examples}

The graph-theoretic analysis developed in the previous
sections were general and applicable to all single field models of inflation 
employing $\delta N$ formalism. In this section, we study USR inflation as an interesting setup where the previous results can be applied.

%%%%%%%%%%%%%%%%%%%%%%%%%%%%%%%%%%%%
\subsection{USR Inflation with Transition to SR}

The USR setup is a single field model of inflation in which the potential is exactly flat $V=V_0$ \cite{Kinney:2005vj, Namjoo:2012aa, Martin:2012pe, Morse:2018kda, Dimopoulos:2017ged}. As a result, the first SR parameter scales like $\epsilon \propto a(t)^{-6}$ while the second SR parameter is large, $\eta= -6$. Since $\epsilon$ falls off exponentially, the curvature perturbation grows on superhorizon scales like $R \propto a(t)^{3}$. This is the key reason why the USR setup violates the Maldacena consistency condition, predicting 
$f_{NL}=5/2$ \cite{Namjoo:2012aa}.  Since the curvature perturbations grow exponentially during the USR phase, the system becomes non-perturbative if the duration of USR phase is unlimited. Therefore, one has to terminate the USR phase after which the system enters a SR phase. The transition from the non-attractor USR phase to the final attractor SR phase can be either sharp or mild. The original USR model studied in \cite{Namjoo:2012aa, Chen:2013aj} was an idealized setup in which the transition from a USR phase to the attractor phase is infinitely sharp so the system enters the attractor phase immediately after the transition. However, in a realistic setup one expects that the transition from the non-attractor phase to the attractor phase to take some time. This indeed is controlled by a dimensionless parameter, $h$, which is defined as  \cite{Cai:2018dkf},
\ba
\label{h-def}
h\equiv  \frac{6 \sqrt{2 \epsilon_V}}{\dot \phi(t_e)} M_P H= 
-6 \sqrt{\frac{\epsilon_V}{\epsilon_e}} \, ,
\ea
in which $\epsilon_e$ is the value of $\epsilon$ at the end of USR phase $t=t_e$ while $\epsilon_V$ is its value during the final attractor phase. Here, without loss of generality, we have assumed that $\dot \phi <0$ so $h<0$. 
The infinitely sharp transition considered in \cite{Namjoo:2012aa, Chen:2013aj} corresponds to $h \rightarrow -\infty$.

During the USR phase the system is in a non-attractor phase so the evolution of the system should be considered in the full phase space in which $N=N(\phi, \dot \phi)$. This is the important difference between $\delta N$ formalism in USR setup compared to the standard SR setups in which the system is in the attractor phase and $N= N(\phi)$. 

In this subsection we consider the setup in which the potential is continuous across the transition from USR to the SR phase but its derivative has a jump. In the next subsection we relax the first assumption as well in which the potential itself has a jump (step) across the transition.   

The total number of e-fold $N(\phi, \dot \phi)$ was calculated in  \cite{Cai:2018dkf}, yielding,
\ba
\label{N-tot}
N_{\text{tot}} = N_{\text{USR}} + N_{\mathrm SR}= 
-\frac{1}{3} \ln \Big[1+ \frac{3 (\phi - \phi_e)}{\pi}\Big]+ \frac{1}{\eta_V} \ln\Big[ - 2 \eta_V \pi_e - 6 \sqrt{2 \epsilon_V} M_P
\Big] +  \text{constant} \, ,
\ea
in which $N_{\text{USR}}$ and $N_{\mathrm SR}$ are the contributions in total number of e-folds during the USR phase and the second SR phase respectively while the constant piece is independent of the initial values in phase space.   Furthermore, $\pi\equiv \phi'(N)= \sqrt{2 \epsilon} M_P$ and  
$(\phi_e, \pi_e)$ are the values of the field and its momentum  at the end of USR. Also, $\eta_V \ll 1$ is the value of the second SR parameter (related to second derivative of $V(\phi) $)  during the final attractor phase.  In addition, the following constraint between $(\pi_e, \phi_e)$ and the initial values $(\phi, \pi)$ holds,
\ba
\label{constr}
\pi_e= \pi+  3 (\phi- \phi_e) \, .
\ea

When employing $\delta N$ formalism, it is important to realize that the variations are with respect to initial values, $(\phi, \pi)$ while the position of the end of USR is fixed in the field space at $\phi=\phi_e$. On the other hand, $\pi_e$ is not fixed in phase space and is related to the initial values $(\phi, \pi)$ and $\phi_e$ via Eq. (\ref{constr}). Finally, after taking the variation, only  $\delta \phi$ is relevant in $\delta N$ expansion since 
$\delta \pi$ is negligible as it approaches zero for light fields on superhorizon scales.

We are interested in the limit $\eta_V\to0$ in which the final phase of inflation is in the attractor regime as in conventional SR models. Taking the variation of Eq.~\eqref{N-tot} with respect to the initial field value,  we obtain
\ba
\delta N = \frac{6}{h}\frac{\delta\phi}{\pi_e} -\frac13\ln\Big( 1+\frac{3\delta\phi}{\pi_e} \Big).
\label{USR-deltaN}
\ea
Note that  $h$ and $\pi_e$ depends on the initial data in the phase space. 
In particular, $\pi_e$ is related to the initial data via the constraint Eq. (\ref{constr}).

Expanding $\delta N$ in Eq. (\ref{USR-deltaN})  formally in powers of $\delta\phi$ via, 
\begin{equation}
\delta N =\sum_{d=1}^{\infty} \frac{N^{(d)}}{d!}(\delta\phi)^d,
\end{equation}
we find,
\ba
N'= \frac{-1}{\gamma \pi_e} , \qquad
N^{(d)} = \frac{(-1)^d3^{d-1}(d-1)!}{\pi_e^d}, \qquad d\ge2 \, ,
\label{usr-derivatives}
\ea
where for convenience we have defined,
\ba
\gamma\equiv\frac{h}{h-6},  \qquad \qquad 0<\gamma<1 \, .  
\label{usr-gamma}
\ea
Using the definition of normalized vertex amplitudes $a_d$ introduced in Section \ref{amplitudes-sec}, the above expressions yield, 
\ba
a_1 =1 \, ,\quad \quad 
a_d= \frac{N^{(d)}}{(N')^d} = 3^{d-1}(d-1)!\gamma^d \, , \qquad \qquad 
 d \ge 2 \, .
\label{usr-ad-all}
\ea
Since $\gamma>0$, an immediate conclusion from the above result is 
that $a_d>0$ for all $d\ge2$.

The above expression determines the complete hierarchy of normalized vertices for the USR to SR transition. In particular,  the amplitude of bispectrum is obtained to be, 
\ba
F_{\rm NL}=a_2 =\frac{3h^2}{(h-6)^2} = 3 \gamma^2 \, .
\label{usr-fnl}
\ea
We realize that the maximum value of $F_{NL}$ is obtained in the limit of an infinitely sharp transition $h\rightarrow -\infty$.  Taking the conventional factor $5/6$ into account, in this limit we obtain $f_{NL}= 5 F_{NL}/6=5/2$ as obtained originally in \cite{Namjoo:2012aa}. On the other hand, in mild transitions with $|h| \lesssim 1$, much of the non-Gaussianity accumulated during the USR phase is washed out during the subsequent attractor phase as observed  in\cite{Cai:2018dkf}.

We can now apply the graph-theoretic results of the previous sections.
As we have seen,  for a tree $T$ with $n$ vertices and vertex degrees $d_v$, the  amplitude $f_T$ of the tree is given by the products of $a_{d_v}$  as given in  Eq. (\ref{fT-ad}). 
%\begin{equation} f_T=\prod_{v\in V(T)}a_{d_v}. \label{usr-tree-product} \end{equation}
Now, let $L(T)$ denotes the number of leaves of the tree. Since the leaves have $a_1=1$, using Eq.~\eqref{usr-ad-all} and the tree identities,
\begin{equation}
\sum_v d_v=2(n-1),
\qquad
\sum_{v:\,d_v\ge2}(d_v-1)=n-2,
\end{equation}
the amplitude $f_T$ from  Eq. (\ref{fT-ad}) is obtained to be, 
\ba
f_T = 3^{n-2} \gamma^{\,2n-2-L(T)} \prod_{v\in V(T)}(d_v-1)!  \, .
%\nonumber\\
%&= & F_{NL}^{n-2} \gamma^{- L(T)} \prod_{v\in V(T)}(d_v-1)! \, .
\label{usr-tree-all}
\ea
This is the general result for the amplitude of an arbitrary
tree in the USR to SR model. An interesting consequence is that the transition parameter affects different
tree topologies differently. At a fixed number of vertices $n$, the power of
$\gamma$ depends on the number of leaves, while the remaining combinatorial factor depends on the vertex degrees. Finally, we see that the amplitudes are positive for all shapes. This is a strong and interesting result, extending the result of \cite{Firouzjahi:2023xke} established for the particular case $d_v=2$. 

As an example, consider the trispectrum $n=4$. For the star tree with the 
degree sequence $(3,1,1,1)$ we obtain,
\ba
f_{(3,1,1,1)}= G_{NL}= \frac{18 h^3}{ (h-6)^3} \, .
\ea
On the other hand, for the path tree with the degree sequence $(2,2,1,1)$ corresponding to $T_{NL}$, we obtain
\ba
f_{(2,2,1,1)}= \frac{9 h^4}{(h-6)^4} \, .
\ea
Both of the above results  agree with the results of \cite{Firouzjahi:2025ipu}. 

As the second example, let us compare the amplitudes of the star graph $S_n$ and path graph $P_n$ for a general value of $n$.  The star graph  contains $n-1$ leaves and one vertex of degree $n-1$, so  its amplitude is,
\ba
f_{S_n} = 3^{n-2}(n-2)!\gamma^{n-1}.
\label{usr-star}
\ea
On the other hand, the path graph $P_n$ contains only two leaves, and its amplitude is, 
\ba
f_{P_n} = 3^{n-2}\gamma^{2n-4}.
\label{usr-path}
\ea
As a result, their ratio is given by, 
\ba
\frac{f_{S_n}}{f_{P_n}} = (n-2)!\gamma^{3-n}.
\label{star-path-ratio}
\ea
For a mild transition, $\gamma\ll 1$, the path amplitude is more strongly suppressed than the star amplitude. The difference becomes increasingly pronounced at higher orders. On the other hand, in the infinitely sharp transition limit, $\gamma\to1$, the relative amplitudes reduce to purely combinatorial factors.

More generally, for any two trees $T_1$ and $T_2$ with the same number of vertices $n$, Eq.~\eqref{usr-tree-all} gives
\ba
\frac{f_{T_1}}{f_{T_2}} = \gamma^{L(T_2)-L(T_1)}\, ~ 
\frac{\displaystyle\prod_{v\in V(T_1)}(d_v-1)!}
{\displaystyle\prod_{w\in V(T_2)}(d_w-1)!}.
\label{usr-ratio}
\ea
In particular, we see that trees with the same degree sequence have identical amplitudes even when their momentum shapes are different. 
As we have observed in previous Section,  this
phenomenon first occurs in the six-point functions.

The factorization identities derived in Section \ref{amplitudes-sec} also hold for this model. For example, one can check that all the factorization products listed in Table \ref{prim-amp-table} apply here as well. For example, for the five-point amplitudes we have the following relations,  $f_{(3,2,1,1,1)}
=a_3a_2$ and $f_{(2,2,2,1,1)} =a_2^3$ and so on for higher order correlations.  These relations follow directly from the graph structure and do not require additional dynamical assumptions. The USR dynamics enters through the specific values of the vertices $a_d$ in Eq.~\eqref{usr-ad-all}.

\subsection{USR Transition with a Potential Step}
\label{subsec:usr-step}

We now extend the preceding analysis to the setup in which the potential has a jump $\Delta V$ across the transition. Following the analysis of
Ref.~\cite{Cai:2022erk}, the transition is characterized
by two parameters, $h$ and $g$ in which $h$ as before measures the sharpness of the transition while $g$ measures the discontinuity in the potential across the transition.  As we shall see, the additional parameter $g$ changes the hierarchy of nonlinear vertices and introduces the possibility of cancellations among different contributions. These cancellations have direct consequences for the tree amplitudes and their momentum shapes.

Let $\pi_e$ and $\pi_d$ denote the inflaton velocities immediately before and after the potential step, respectively. As in previous example, we assume $\pi <0$  and  the constraint Eq. (\ref{constr}) does hold. 
Assuming that the Hubble parameter is approximately constant across the transition, from the energy conservation we obtain \cite{Cai:2022erk}, 
\begin{equation}
%\pi_c=\pi+3(\phi-\phi_e),\qquad
\pi_d=-\sqrt{\pi_e^2-\frac{6M_{\rm Pl}^2\Delta V}{V_0}},
\label{eq:step-velocities}
\end{equation}
where $\Delta V$ is the change in potential energy across the step. Let us  define,
\begin{equation}
g\equiv\frac{\pi_d}{\pi_e}>0,\qquad
h\equiv
\frac{6\sqrt{2\epsilon_V}M_{\rm Pl}}{\pi_d}<0.
\label{eq:step-gh}
\end{equation}
 An upward step with  $\Delta V>0$ corresponds to $0<g<1$, whereas a downward step with  $\Delta V<0$ gives $g>1$. The case $g=1$ recovers the transition studied in previous subsection. 

As before, the total number of e-folds contains the USR and SR contributions which is given by \cite{Cai:2022erk}, 
\begin{equation}
N(\phi,\pi)=
-\frac13\ln\left(\frac{\pi_e}{\pi}\right) +\frac{1}{\eta_V}
\ln\left( -2\eta_V\pi_d -6\sqrt{2\epsilon_V}M_{\rm Pl} \right) +\mathrm{constant}.
\label{eq:step-N}
\end{equation}
In calculating $\delta N$ note that we take the variations with respect to
$\delta \phi$ and $\delta \pi$ while holding the point of transition $\phi=\phi_e$ and the point of end of inflation $\phi=\phi_f$ fixed.  In particular, the parameters $g$ and $h$ are evaluated on the fiducial background only after performing these variations.

As in previous example, we are interested in the limit where  
$\eta_V\to0$ so the second phase is in the SR regime. Neglecting
the initial momentum fluctuation as before, the resulting curvature perturbation is given by, 
\ba
\begin{split}
\mathcal R=\delta N =-\frac13\ln\left( 1+\frac{3\delta\phi}{\pi_e} \right)
+\frac{2}{h} \left[ \sqrt{ 1+\frac{6\delta\phi}{g^2\pi_e} +\frac{9(\delta\phi)^2}{g^2\pi_e^2} } -1 \right]. \end{split}
\label{eq:step-deltaN}
\ea
The logarithmic contribution above originates from the USR evolution whereas the square root contribution reflects the change in velocity  across the potential step and the subsequent SR evolution.

Expanding Eq.~\eqref{eq:step-deltaN} in powers of $\delta\phi$ and calculating the normalized vertex amplitudes $a_d$ as in  Eq. (\ref{ad-def}), for the amplitudes relevant to $n=6$, we obtain, 
\ba
\begin{aligned}
a_2&=
\frac{3h[g^4h+6(g^2-1)]} {(g^2h-6)^2},\\[2mm]
a_3&=
\frac{18h^2[g^6h+9(g^2-1)]} {(g^2h-6)^3},\\[2mm]
a_4&=
\frac{162h^3[g^8h-3(g^2-1)(g^2-5)]} {(g^2h-6)^4},\\[2mm]
a_5&=
\frac{486h^4[4g^{10}h -15(g^2-1)(3g^2-7)]} {(g^2h-6)^5}.
\end{aligned}
\label{step-vertices}
\ea
On can check that for $g=1$, the above results reduce to  $a_d$ in Eq. (\ref{usr-ad-all}). 

%The general graph-theoretic rule remains unchanged. The amplitude $f_T$, as in Eq. (\ref{fT-ad}), is given by the products of  $a_{d}$.  The new physics therefore enters through the dependence of the vertices amplitudes $a_d$ on the parameters $(h, g)$.

From the above results for $a_d$ we see that there can exist situations in which some amplitudes $a_d$ vanish. This corresponds to non-trivial competition between the parameters $(h, g)$. In our setup $h<0$, but  depending on whether the jump in potential is upward (downward) then   $0< g<1$ ($g >1$).  So let us consider these two cases in turn. 

%%%%%%%%%%%%%%%%%%%%%%%%%%%%%%%%%%%

\subsubsection{Positivity of Amplitudes for $0<g<1$}

We first consider an upward potential step, for which the
inflaton loses kinetic energy when crossing the discontinuity,
corresponding to $0<g<1$. As can be checked explicitly from
Eq. (\ref{step-vertices}), the nonlinear vertices $a_2,\ldots,a_5$ are
positive in this case 
so there is no cancellation in the  amplitudes.  We have confirmed in Appendix \ref{sign-app} that this conclusion does hold for all values of $a_d$. Therefore, 
\ba
a_d>0,\qquad \qquad 
( 0<g<1) \, .
\ea

An immediate consequence follows from the graph-theoretic
factorization relation Eq. (\ref{fT-ad}). For any connected tree $T$, its  amplitude is positive for $0< g \leq1$, 
\ba
f_T=\prod_{v\in V(T)}a_{d_v}>0 \, , \qquad 
\qquad (0<g<1) \, .
\ea
This is a strong and interesting result.  An upward potential step cannot generate a vanishing tree amplitude through the cancellation of nonlinear vertices at any order. The parameter $g$
modifies the relative magnitudes of different tree amplitudes, but not their signs.

%%%%%%%%%%%%%%%%%%%%%%%%%%%%%%%%%%%

\subsubsection{Cancellation of Amplitudes for $g>1$}

Now consider the case of a downward  potential step, $g>1$. In this case the inflaton gains kinetic energy at the discontinuity, and the nonlinear
vertices can receive contributions with opposite signs.
Unlike the upward step case, the amplitudes $a_d$ can take either signs and in some situations they can vanish. This effect was observed in \cite{Namjoo:2024ufv} for bispectrum 
and here we extend this study to higher orders. The details of the analysis where the cancellation of the amplitudes occur in the parameter space  of $(h, g)$ are presented in Appendix \ref{sign-app}.

The first two cancellation curves follow immediately
from Eq. (\ref{step-vertices}), yielding, 
\ba
\begin{aligned}
a_2=0:\qquad
&h=-\frac{6(g^2-1)}{g^4}\, , \\
a_3=0:\qquad
&h=-\frac{9(g^2-1)}{g^6}.
\end{aligned}
\ea
Both lie in the region $h<0$ for $g>1$. They intersect at,
\ba
g=\sqrt{\frac32}\, , \qquad \qquad h=-\frac43,
\ea
where $a_2=a_3=0$. At this point, the tree level bispectrum and both of trispectrum amplitudes vanish ($F_{NL}= G_{NL}= T_{NL}=0$), while higher order non Gaussianity can survive. This is a unique situation that can occur in the downward step transition. Of course, there can be subleading SR contributions in bispectrum and trispectrum which are not captured by our leading $\delta N$ analysis. But their magnitudes are considerably small and may not be detectable observationally. 

%Importantly, these cancellations do not require the linear response to vanish. Indeed,
%\ba N'=\frac{6-g^2h}{hg^2\pi_c}, \ea
%which remains nonzero for finite $h<0$ and $g>1$. Thus a vanishing nonlinear vertex does not imply a vanishing curvature power spectrum.

The graph-theoretic interpretation is particularly simple. Since the amplitude of each tree $f_T$ is given by the products of vertex amplitudes $a_d$ as in Eq. (\ref{fT-ad}),  the condition $a_d=0$ eliminates every tree containing at least one vertex of degree $d$. This selection rule
holds at arbitrary order and is independent of the momentum configuration.

As some case examples, consider first the curve $a_2=0$. The bispectrum amplitude vanishes, $F_{\rm NL}=a_2=0$. Similarly,  the path trispectrum amplitude vanishes as well, $f_{(2,2,1,1)}=a_2^2=0$. 
However, the star trispectrum amplitude $f_{(3,1,1,1)}=a_3$ is generally nonzero. Therefore, as expected,  the vanishing of bispectrum does not imply the absence of higher order non-Gaussianity.
On the other hand, at five points, $f_{(4,1,1,1,1)}=a_4, f_{(3,2,1,1,1)}=a_3a_2=0$ and $f_{(2,2,2,1,1)}=a_2^3=0$. As a result,  only the star topology survives generically. Finally, at six points, both the star and double star topologies contain no $a_2$ vertices. Consequently, $f_{(5,1,1,1,1,1)}=a_5, f_{(3,3,1,1,1,1)}=a_3^2$ can survive  while all remaining six-point tree amplitudes vanish on the curve of $a_2=0$.

As the second example, now consider the curve, $a_3=0$. The star
trispectrum vanishes, whereas the path trispectrum generally survives, $f_{(3,1,1,1)}=0, f_{(2,2,1,1)}=a_2^2$. At five points, the mixed topology
$(3,2,1,1,1)$ disappears, while the star and path topologies generally remain. At six points, the two distinct topologies with degree sequence
$(3,2,2,1,1,1)$ vanish simultaneously, despite their different momentum shapes. The double-star topology $(3,3,1,1,1,1)$ also vanishes. The remaining amplitudes are proportional to $a_5$, $a_4a_2$ and $a_2^4$.

At the intersection $a_2=a_3=0$, these selection rules combine. All three- and four-point tree amplitudes vanish, whereas the five- and six-point
star amplitudes $a_4$ and $a_5$ remain nonzero. This illustrates how lower-order non-Gaussianity can be suppressed while higher order correlations are retained.

We conclude that downward potential steps admit a hierarchy of possible vertex zero curves. Whenever a vertex vanishes, entire families
of tree topologies are eliminated according to their degree sequences. This provides a direct connection between the inflationary transition
parameters and the graph theoretic organization of higher order correlation functions. Having said this, we comment that we have idealized the setup by an instant and sharp transition. One may consider a relaxed situation where the transition is smooth. Whether or not the cancellation of some topologies occur needs to be examined in these realistic situations.

%%%%%%%%%%%%%%%%%%%%%%%%%%%%%%%%%%%%%%

\section{Summary and Discussions}
\label{summary}

In this work we studied the $\delta N$ formalism in general
single field models of inflation and investigated the
$n$-point correlation functions using graph theory methods.
A great advantage of the $\delta N$ formalism is that one
only needs to follow the background trajectory
$N=N(\phi,\dot\phi)$ in phase space. The curvature
perturbation on superhorizon scales is related to the
perturbation in the number of e-folds, which can be
expanded perturbatively in terms of the initial field
fluctuations. This provides a natural starting point
for organizing higher order correlation functions using
established results from graph theory.

Using graph theory, we studied connected tree diagrams
and classified their independent, non-isomorphic topologies. Each vertex is characterized by its degree, and the corresponding normalized vertex amplitude is $a_d=N^{(d)}/(N')^d$, with $a_1=1$. We showed that
the amplitude of an arbitrary tree $T$ factorizes as $f_T=\prod_{v\in V(T)}a_{d_v}$. This result is independent of the details of the
inflationary model and allows the amplitudes of
higher order diagrams to be constructed directly
from their vertex degrees. In particular, the star
diagrams play the role of primary diagrams, while
the amplitudes of other trees are the products of the
corresponding primary amplitudes. The factorization
also provides a simple graph-theoretic interpretation
of the familiar Suyama-Yamaguchi  equality
at tree level and its higher order counterparts. The relations between amplitudes are algebraic consequences of the vertex structure, whereas the corresponding relations between complete momentum channels require the appropriate sums over external momentum assignments.

An important distinction is that the degree sequence determines the  amplitude but does not necessarily determine the topology of a tree. Two non-isomorphic trees may have the same degree
sequence and therefore the same amplitude, while their momentum dependent shapes are different. The number of non-isomorphic trees is governed by
the well-known Otter counting problem and increases
rapidly with the number of vertices. The graph
representation therefore becomes particularly useful
at higher orders, where a direct calculation of all
momentum permutations becomes increasingly cumbersome.
It also makes clear which properties of the
correlation functions are universal and which
depend on the underlying inflationary model.

To illustrate the applications of this framework, we considered a transient USR phase followed by a SR attractor. The sharpness of the transition from the USR phase to the final attractor regime is controlled by the parameter $h <0$.  For a transition without a potential step $(g=1)$, the nonlinear vertex amplitudes can be obtained analytically at arbitrary order as given in Eq. (\ref{usr-ad-all}). This in turn gives  the amplitude of every connected tree diagram at arbitrary order as given in Eq. (\ref{usr-tree-all}). We have shown that the transition parameter affects different
topologies differently, so that the relative strengths of the higher point correlation functions depend on the details of the USR to SR transition. An important conclusion was that all amplitudes are positive.  

%%%%%%%%%

We then generalized the analysis by including a potential step, introducing the additional parameter \(g\) which measures the change in the inflaton velocity across the transition. Both upward steps, \(0<g<1\), and downward steps, \(g>1\), are allowed. An important result of our analysis is that, for upward steps, all nonlinear vertex amplitudes \(a_d\) are strictly positive at arbitrary order. Consequently, all connected tree amplitudes are positive, and no cancellation of individual tree amplitudes can occur.  For downward steps, the situation is qualitatively different. The competition between the USR evolution and the velocity change across the step allows the nonlinear vertex amplitudes to change sign or vanish. 
Their graph-theoretic interpretation is particularly simple: whenever \(a_d=0\), every tree containing a vertex of degree \(d\) has vanishing amplitude, while trees without such vertices may survive. Thus, the cancellation of a single nonlinear response coefficient eliminates entire families of higher-point tree topologies. As explicit examples, we studied the cancellation curves \(a_2=0\) and \(a_3=0\) and examined their consequences for five- and six-point correlation functions. These results demonstrate that lower order non-Gaussianity can be suppressed without necessarily eliminating higher order correlations. More generally, the pattern of surviving and vanishing tree amplitudes provides information about the underlying inflationary transition that cannot be inferred from the bispectrum and trispectrum alone.

There are several directions in which the current study can be extended. A natural extension is to include loop diagrams and investigate how the resulting structures are related to corresponding theorems in graph theory. In particular, it would be interesting to examine which tree-level
factorization relations and SY-type equalities survive after loop corrections
are included. Another direction is to consider multifield models of inflation.
In such scenarios, the graph structure becomes considerably richer because
vertices and edges may carry different field labels, and the single field SY
equality is replaced, in general, by the SY inequality. Finally, to perform the analytical calculations, here we considered an instantaneous and sharp transition from USR to SR phase. It would be desirable to extend the current analysis to the realistic setups with smooth transition and examine whether or not the current conclusions about the sign and the amplitude cancelation hold.

\vspace{0.7cm}

{\bf Acknowledgments:}  
We thank Amjad Ashoorioon, Mohammad Hossein Namjoo and Farnaz Bahmani  for useful discussions and correspondences. The author acknowledges the use of Maple and Mathematica software packages for analytical and numerical analysis.  The work of A. N. is partially supported by the Iran National Science Foundation (INSF).
The work of H. F.  is supported by the INSF  of Iran under the grant  number 4045105. \\

{\bf Use of artificial intelligence:} During the preparation of this manuscript, the authors used  ChatGPT (GPT-6) to assist with the development of figures, mathematical analysis, manuscript organization and editing.  The research questions and underlying theoretical framework were developed by the authors. All AI-assisted derivations, figures, and interpretations were independently verified and critically examined by the authors, who take full responsibility for this work.

%%%%%%%%%%%%%%%%
\appendix
\section{Signs of Amplitudes}
\label{sign-app}

In this Appendix we study the signs of the vertex amplitudes $a_d$.
We show that for the upward step with $0< g<1$ all $a_d$ are positive while for the downward step with $g>1$ there are corners of parameter space $(h, g)$ where $a_d$ can vanish. 
%%%%%%%%%%%%%%%%%%%%%%%%%%%%%%%%%%%

Let us introduce the dimensionless variable $x\equiv -3\delta\phi/\pi_e$. Equation~(6.23) then takes the following form,
\ba
\delta N(x)=-\frac{1}{3}\ln(1-x) +\frac{2}{h}\Bigg(
\sqrt{1-\frac{2x}{g^2}+\frac{x^2}{g^2}}-1 \Bigg).
\ea
For convenience, let us define the following Taylor expansion,
\ba
\label{f(t,z)}
\sqrt{1-2tz+z^2}
=\sum_{d=0}^{\infty}c_d(t)z^d,
\quad \qquad t=\frac{1}{g}, \quad \qquad z=\frac{x}{g} \, ,
\ea
where $c_d(t)$ is the coefficients of Taylor expansion. 
Using the generating function of Legendre polynomials \cite{Arfken},
the coefficients for $d\geq2$ can be expressed as,
\ba
c_d(t)=P_d(t)-2tP_{d-1}(t)+P_{d-2}(t).
\ea
Expanding $\delta N$ and using
$N'=-\pi_e^{-1}(1-6/hg^2)$, we obtain the vertex amplitude as follows, 
\ba
\label{ad-sign}
a_d= \frac{3^{d-1}(d-1)!} {\left(1-\frac{6}{hg^2}\right)^d}
\left[ 1+\frac{6d}{hg^d} c_d\big(\frac{1}{g}\big) \right], \qquad d\geq2. 
\ea
This expression reproduces the explicit results in
Eq. (\ref{step-vertices}). In particular, for $g=1$ it reduces to our previous result Eq. (\ref{usr-ad-all}).

%%%%%%%%%%%%%%%%%%%%%%%%%%%%%%%%%%
\subsection{Upward Step ($0<g<1$)}
\label{upward-app}

The above analysis were general. We now establish the sign of $c_d(t)$ in the case of upward step, corresponding to $0<g<1$ and
$t>1$.  Differentiating the generating function twice gives,
\ba
\label{f-second}
\frac{d^2}{dz^2}\sqrt{1-2tz+z^2}
=(1-t^2)(1-2tz+z^2)^{-3/2}.
\ea
For  $t>1$ the quadratic polynomial factorizes as,
\ba
1-2tz+z^2
=(1-\lambda_+z)(1-\lambda_-z), \qquad
\lambda_\pm=t\pm\sqrt{t^2-1}>0\, .
\label{lambda-pm}
\ea
Both factors $(1-\lambda_\pm z)^{-3/2}$ have strictly
positive Taylor coefficients. Correspondingly, their product
has strictly positive coefficients as well. Since $1-t^2<0$, this establishes that the Taylor coefficients of the left hand side of Eq. (\ref{f(t,z)}) for $z^m, m\ge2$ are all negative.  Consequently,  comparison of the right hand side of Eq. (\ref{f(t,z)}) yields,
\ba
c_d(t)<0,\qquad \qquad  (d\geq2,\quad t>1).
\ea
In conclusion, for $h<0$, the correction proportional to $c_d(1/g)$ in the square brackets of the general expression for $a_d$ in Eq. (\ref{ad-sign}) 
is positive. Moreover, $1-6/(hg^2)>0$ so we conclude the
all order positivity relation,
\ba
a_d>0,\qquad \qquad 
(d\geq2,\quad 0<g\leq1) \, .
\ea
Note that the limiting case $g=1$ follows directly from Eq. (\ref{usr-ad-all}).

Using the graph-theoretic factorization relation Eq. (\ref{fT-ad}), the above results indicate that for any connected tree $T$, its  amplitude is positive, 
\ba
f_T=\prod_{v\in V(T)}a_{d_v}>0 \, , \qquad 
\qquad (0<g\leq1) \, .
\ea

%%%%%%%%%%%%%%%%%%%%%%%%%%%%%%%%%%
\subsection{Downward Step ($g>1$)  }
\label{downward-app}

Now consider the downward step with $g>1$. Following the same step as in previous section, the Taylor expansion 
Eq. (\ref{f(t,z)}) and the formula for $a_d$ as in Eq. (\ref{ad-sign}) are still valid. However, since now $t= 1/g<1$, then the characteristics $\lambda_\pm$ become complex and the arguments following Eq. (\ref{lambda-pm}) do not proceed. As a result, $c_d(1/g)$ can take either signs which, depending on the model parameters, may yield to vanishing $a_d$. 

To handle this properly, note that the coefficients $c_d$ can be expressed in terms of Gegenbauer polynomials. Indeed, using the second derivative 
given in Eq. (\ref{f-second})  and the generating function \cite{Arfken}, 
\ba
(1-2tz+z^2)^{-3/2}
=\sum_{m=0}^{\infty}C_m^{(3/2)}(t)z^m\, ,
\ea
we obtain,  
\ba
c_d(t)=\frac{1-t^2}{d(d-1)}
C_{d-2}^{(3/2)}(t),\qquad d\geq2 \, ,
\ea
where $C_{d-2}^{(\alpha)}(t)$ are the ultraspherical polynomials (proportional to the Gegenbauer polynomials). 

Substituting this value of $c_d(1/g)$ into the general expression
for $a_d$  in Eq. (\ref{ad-sign}), we obtain the requirement for the 
cancellation of the vertex amplitude as a condition on the value of $h$, 
\ba
a_d=0
\quad\Longleftrightarrow\quad
h=-\frac{6(g^2-1)}{(d-1)g^{d+2}}\,
C_{d-2}^{(3/2)}\big(\frac{1}{g}\big),
\qquad \qquad  (d\geq2) \, .
\ea
Since $h<0$ and $g>1$, a physically admissible cancellation curve exists where, 
\ba
C_{d-2}^{(3/2)}\big(\frac{1}{g}\big)>0 \, .
\ea
This provides a general criterion for the existence of vertex zero curves at arbitrary order. Indeed, one can examine that the Gegenbauer coefficient
 $C_{d-2}^{(3/2)}(t)$ has a root only for $t<1$ so the cancellation of vertex amplitude can occur for $g >1$ while it is not possible for $g<1$.

%%%%%%%%%%%%%%%%

\bibliography{deltaN-graph}{}

\bibliographystyle{JHEP}
%\bibliographystyle{JHEPNoTitle}
%%%%%%%%%%%%%%%%%%%%%%%%%%%%%%%%%%%%%%%%%%%%%%%%%%%%%

\end{document}